\documentclass[aps,prd,onecolumn,nofootinbib,superscriptaddress]{revtex4}
\pdfoutput=1
\usepackage{float}
\usepackage{graphicx}
\usepackage{amsmath}
\usepackage{amsfonts}
\usepackage{amssymb,ulem}
\usepackage{color}%
\usepackage{dcolumn}
\usepackage{subfigure}
\usepackage{multirow}

\usepackage{MnSymbol,wasysym}
\usepackage{braket,diagbox}
\usepackage{eurosym}
\usepackage{calrsfs}
\usepackage[usenames,dvipsnames,svgnames]{xcolor}

\newcommand{\RNum}[1]{\uppercase\expandafter{\romannumeral #1\relax}}
\usepackage[colorlinks=true,linkcolor=blue,urlcolor=blue,filecolor=black,citecolor=red,
pdfstartview=FitV,pdftitle={},pdfsubject={},pdfkeywords={},pdfpagemode=None,bookmarksopen=true]{hyperref}

\usepackage{pifont}

\begin{document}
\baselineskip=0.4 cm

\title{Effects of hair on the image of a rotating black hole illuminated by a thin accretion disk}

\author{Yuan Meng}
\email{mengyuanphy@163.com}
\affiliation{Center for Gravitation and Cosmology, College of Physical Science and Technology, Yangzhou University, Yangzhou, 225009, China}

\author{Xi-Jing Wang}
\email{xijingwang01@163.com}
\affiliation{Department of Astronomy, School of Physics and Technology, Wuhan University, Wuhan 430072, China}

\author{Yong-Zhuang Li}
\email{liyongzhuang@just.edu.cn}
\affiliation{Research center for theoretical physics, School of Science, Jiangsu University of Science and Technology, Zhenjiang 212100, China.}

\author{Xiao-Mei Kuang}
\email{xmeikuang@yzu.edu.cn (corresponding author)}
\affiliation{Center for Gravitation and Cosmology, College of Physical Science and Technology, Yangzhou University, Yangzhou, 225009, China}

\date{\today}

\begin{abstract}
\baselineskip=0.5 cm
In this paper, we investigate the shadow and optical appearance of the hairy Kerr black hole illuminated by a thin accretion disk, the materials of which outside the innermost stable circular orbit (ISCO) move on the equatorial circular orbit, while inside the ISCO they quickly plunge into the black hole. The deformation parameter $\alpha$ and hair parameter $l_o$ are found to influence the motions of accretion as well as the redshift effect of the photon, such that they  significantly affect the shadow and image of the hairy Kerr black hole. Especially, these two parameters have competing effects on the size of the black hole's shadow, and significantly increase the width of photon ring. This study provides a preliminary theoretical prediction that the image of the hairy Kerr black hole, especially the photon ring structure, may be used to constrain the hair parameters with future high-precision astronomical observation.
\end{abstract}


\maketitle
\tableofcontents

\newpage
\section{Introduction}

Recently, the Event Horizon Telescope (EHT) has achieved  remarkable  breakthroughs, offering new insights into the physical properties of strong gravitational fields. The EHT successfully captured images of two supermassive black holes: M87* \cite{EventHorizonTelescope:2019dse,EventHorizonTelescope:2019uob,EventHorizonTelescope:2019jan,EventHorizonTelescope:2019ths,EventHorizonTelescope:2019pgp,EventHorizonTelescope:2019ggy} and Sgr A* \cite{EventHorizonTelescope:2022wkp,EventHorizonTelescope:2022apq,EventHorizonTelescope:2022wok,EventHorizonTelescope:2022exc,EventHorizonTelescope:2022urf,EventHorizonTelescope:2022xqj}. These images represent a significant advancement in astronomical observation, providing an unprecedented opportunity to test general relativity (GR) under extreme physical conditions.
The central dark region in these images, known as the black hole shadow, results from strong gravitational lensing that prevents light from escaping the black hole \cite{Virbhadra:1999nm,Virbhadra:2002ju,Virbhadra:2007kw,Synge:1966okc,Bardeen:1972fi,Bozza:2010xqn}. The outline of the shadow corresponds to the border of the regions that the photons either escape to infinity or are captured  by the black hole.  Surrounding the shadow, the bright ring-like structure, referred to as the photon ring, is formed by light rays emitted from matters within the accretion flow. This material, mainly composed of gas and dust, orbits the black hole at high velocities, generating enormous amounts of energy as it spirals inward under the influence of the black hole's strong gravity. The light emitted by the accretion flow is bent  by the strong gravitational field, creating the luminous ring observed in the images. This phenomena provides invaluable experimental evidence for investigating black hole accretion processes, matter dynamics, magnetic fields, and so on. Moreover, it also provides critical insights into the fundamental physical laws governing the most extreme environments in the universe.

The study of black hole shadows originated from investigations into the photon escape cone \cite{Synge:1966okc}, which corresponds to the critical impact parameter of the photon sphere. Early works also explored the angular radius of the photon capture region for a Schwarzschild black hole \cite{Luminet:1979nyg}, while Bardeen examined the distinctive D-shaped shadow of a Kerr black hole \cite{Bardeen:1972fi}, arising from the Lense-Thirring effect  induced by the black hole's rotation. These foundational studies spurred significant interest in black hole shadow research, leading to extensive numerical simulations of shadows across a variety of black holes \cite{Shen:2005cw,Yumoto:2012kz,Atamurotov:2013sca,Papnoi:2014aaa,Kumar:2018ple}. These investigations revealed that black hole shadows are highly sensitive to the spacetime geometry in which they are embedded. In addition, black hole shadows have been extensively studied in the context of alternative gravitational theories and higher-dimensional spacetimes, and even a naked singularity can produce a shadow resembling that of a black hole \cite{Shaikh:2018lcc,Joshi:2020tlq,Dey:2020bgo}. On the other hand, the black hole images obtained by the EHT are widely interpreted as representations of Kerr black holes in GR. However, the finite resolution of the EHT allows for the possibility of alternative explanations, opening a window to explore modified theories of gravity. This has led to considerable interest in using EHT observations to test these theories and constrain their parameters, see for examples \cite{Vagnozzi:2022moj,Afrin:2021imp,Kuang:2022ojj,Tang:2022hsu,Tang:2022bcm,Kuang:2022xjp,Kumar:2019pjp,Shaikh:2021yux,Wu:2023yhp,Capozziello:2023tbo,
Sui:2023rfh,Pantig:2022qak,Ghosh:2023kge,Tsukamoto:2014tja} and references therein.

The image of a black hole is significantly influenced by the accretion material surrounding it, particularly the luminous accretion flow. This matter not only determines the optical characteristics of the black hole but also affects its appearance through complex physical processes. As we described previously, a black hole image is not merely a depiction of spacetime geometry but also a result of the intricate interplay between the surrounding matter and the black hole's extreme gravitational forces. To accurately replicate such images, general relativistic magnetohydrodynamics (GRMHD) simulations are often employed. These simulations model the behavior of electromagnetic plasma in the vicinity of the black hole while accounting for the effects of its strong gravitational field \cite{EventHorizonTelescope:2019pcy}. However, simplified accretion models are often sufficient for capturing the primary features of a black hole image. For instance, early studies using the thin accretion disk model \cite{Luminet:1979nyg,Bambi:2013nla} demonstrated the appearance of primary and secondary images outside the black hole shadow. These images can distinguish between a Schwarzschild black hole and a static wormhole.
Wald et al. further analyzed thin accretion disks around Schwarzschild black holes, identifying three types of emissions: direct emission, lensed ring emission, and photon ring emission \cite{Gralla:2019xty}. While these emissions cumulatively contribute to the observed intensity, the photon ring emission, due to its narrow width, has a negligible impact on the total intensity. Static accretion disks around various black holes have  been extensively studied \cite{Dokuchaev:2019pcx,Peng:2020wun,He:2021htq,Eichhorn:2021iwq,Li:2021riw,Okyay:2021nnh,Wang:2023vcv,Zeng:2023fqy,Hu:2023pyd,Promsiri:2023rez,DeMartino:2023ovj,Sui:2023tje,Zare:2024dtf,Gao:2023mjb,Wang:2024lte} and references therein.
Of particular interest are black holes with two photon spheres, as discussed in \cite{Gan:2021xdl,Gan:2021pwu,Meng:2023htc}. The presence of an additional photon sphere significantly enhances the contribution of photon ring emission to the total observed intensity. Similarly, spherical and infalling spherical accretions have garnered attention, highlighting that the shape and size of the black hole shadow are primarily dictated by the spacetime geometry rather than the specific details of the accretion flow \cite{Zeng:2020dco,Saurabh:2020zqg,Zeng:2020vsj,Qin:2020xzu,Narayan:2019imo}.
Considering more realistic astrophysical scenarios, rotating black holes and their associated accretion images have been a focal point of recent research  \cite{Hou:2022eev,Wang:2023fge,Liu:2021yev,Rodriguez:2024ijx,Zhang:2023bzv,He:2024amh,Zheng:2024brm,Guo:2024mij,Heydari-Fard:2023kgf,Donmez:2024lfi,Yang:2024nin,Li:2024ctu,Wang:2024uda}.

The no-hair theorem states that a general-relativistic black hole is fully and uniquely characterized by its mass, angular momentum, and charge \cite{Ruffini:1971bza}. However, in the presence of additional fields or sources surrounding the black hole, the black hole may acquire additional global charges called ``hairs", which causes its spacetime structure to deviate from the GR description. Recently, the hairy black hole was first constructed based on the ``gravitational decoupling (GD) approach" in \cite{Ovalle:2020kpd,Contreras:2021yxe}. This approach aims to generate black holes with hair by introducing additional sources to describe the deformation of known solutions of GR. Hairy black holes carry additional global charges and could provide more information to test the no-hair theorem. It is worth noting that the hairy black holes obtained by the GD approach exhibit a high degree of generality, because the GD approach does not rely on specific matter fields and allows the use of different types of hair (such as scalar hair, tensor hair, fluidlike dark matter, and so on). Subsequently, the hairy black holes obtained through the GD approach have attracted widespread attention, such as the thermodynamics of rotating hairy black holes \cite{Mahapatra:2022xea}, scalar perturbations and quasinormal modes of hairy black holes \cite{Cavalcanti:2022cga,Yang:2022ifo,Li:2022hkq}, strong gravitational lensing, black hole shadow and image \cite{Afrin:2021imp,Islam:2021dyk}, precession and Lense-Thirring effect \cite{Wu:2023wld} and gravitational waves from extreme mass ratio inspirals \cite{Zi:2023omh}.

Especially, in \cite{Meng:2023htc,Meng:2024puu}, it was found that even though the parameters have significant influences on the rings and shadows of hairy Schwarzschild black hole, its images with certain parameters could be indistinguishable to that of Schwarzschild black hole, namely, the degeneracies of images raise between the hairy Schwarzschild black hole and Schwarzschild black hole. Moreover, the hairy Schwarzschild black hole can have double photon spheres, which introduce new additional rings and accretion features, in contrary to single photon sphere in Schwarzschild black hole in GR. These findings implies that the black hole hair introduced by GD method indeed code significant information on optical features of the hairy static black holes. As an extension of our previous work, this paper aims to investigate the optical appearances of the accretion disk of hairy Kerr black hole. We should elaborately explore the effects of  hairy parameters on the black hole shadows and images of hairy Kerr black hole, and check if the image degeneracies between the hairy Kerr black hole and Kerr black hole can still exist. The answer is not straightforward because different from that around static black hole, the accretion disk surrounding the rotating black hole should move due to the drag force, such that the redshift effects of the photons emitted from disk to the distant observer should be carefully treated.

The paper is organized as follows. In Sec. \ref{sec:background and accretion}, we briefly review the hairy Kerr black hole constructed from GD method, and then analyze the motions of the surrounding accretion disk model. In Sec. \ref{sec-image}, we first review the imaging process by the backward ray tracing method. Then we compare the shadows and images of the hairy Kerr black hole against those of Kerr black hole, and analyze the effects of the hairy parameters. The last section contributes to our conclusion and discussion. All through this paper, we will use the natural unit with $G=c=1$.

\section{Hairy Kerr black hole and its surrounding accretion disk}\label{sec:background and accretion}

In this section, we review the basic ingredients  for the construction of accretion disk images around the hairy Kerr black hole. We begin by specifying the spacetime metric and subsequently outline some well established analytical results related to accretion disk in this background.

\subsection{Hairy Kerr black hole}
As mentioned above, the interactions between black hole spacetimes and matter fields can introduce additional charges, leading to the formation of ``hairs".  These hairs represent new physical effects that modify the black hole's spacetime geometry, resulting in so-called hairy black holes. Recently, Ovalle et.al employed the GD approach to derive a spherically symmetric metric with hair \cite{Ovalle:2020kpd}. In their work, the Einstein equations take the form
\begin{equation}\label{eq-EE}
G_{\mu\nu}\equiv R_{\mu\nu}-\frac{1}{2}Rg_{\mu\nu}=8\pi\Tilde{T}_{\mu\nu},
\end{equation}
where the total energy-momentum tensor $\Tilde{T}_{\mu\nu}$ comprises  two components: $\Tilde{T}_{\mu\nu}=T_{\mu\nu}+\vartheta_{\mu\nu}$. Here  $T_{\mu\nu}$ represents the energy-momentum tensor of a known GR
solution, while $\vartheta_{\mu\nu}$ accounts for new matter fields or a new gravitational sector. The conservation law $\nabla^\mu \Tilde{T}_{\mu\nu}=0$ is satisfied due to the Bianchi identity. The GD approach, originally proposed in  \cite{Ovalle:2017fgl} and extensively utilized in the subsequent work \cite{Contreras:2021yxe,Ovalle:2020kpd}, leverages the decoupling of  $\vartheta_{\mu\nu}$ from  $T_{\mu\nu}$. This framework enables the systematic construction of deformed solutions by separating the equations of motion for the two sectors. To illustrate how the GD approach operates, we consider a spherically symmetric and static solution ${g}_{\mu\nu}$ to the field equations as
\begin{equation}\label{eq-swx0}
ds^2=-e^{\nu(r)}dt^2+e^{\lambda(r)}dr^2+r^2(d\theta^2+\sin^2\theta d\phi^2),
\end{equation}
which generates the Einstein tensor $G_{\mu}^{~\nu}(\nu(r),\lambda(r))$. In this context, the solution can be derived from a seed metric governed solely by the source $T_{\mu\nu}$ (i.e. $\vartheta_{\mu\nu}=0$),
 \begin{equation}\label{eq-swx1}
ds^2=-e^{\xi(r)}dt^2+e^{\mu(r)}dr^2+r^2(d\theta^2+\sin^2\theta d\phi^2).
\end{equation}
The introduction of  $\vartheta_{\mu\nu}$ corresponds to a deformation of the seed metric, expressed as
\begin{gather}\label{eq-deswx}
\xi(r)\rightarrow\nu(r)= \xi(r)+\alpha~k(r), ~~~~~~~e^{-\mu(r)}\rightarrow e^{-\lambda(r)}= e^{-\mu(r)}+\alpha~h(r),
\end{gather}
where the parameter $\alpha$ quantifies the deformation strength. With these deformations, the Einstein equation \eqref{eq-EE} is separated into
\begin{equation}
{G}_{\mu}^{~\nu}(\xi(r),\mu(r))=8\pi T_{\mu}^{~\nu},~~~ \alpha~\mathcal{G}_{\mu}^{~\nu}(\xi(r),\mu(r);k(r),h(r))=8\pi\vartheta_{\mu}^{~\nu}.
\end{equation}
When $\alpha=0$, the tensor $\vartheta_{\mu\nu}$ vanishes, reverting the solution to the original seed metric. The linear decomposition of the Einstein tensor aligns with the linear superposition of sources on the right-hand side of \eqref{eq-EE}, enabling the GD method to function effectively.

To demonstrate the application of this approach, consider the seed metric \eqref{eq-swx1} as the Schwarzschild solution, representing a vacuum case with $T_{\mu\nu}=0$. By introducing an anisotropic fluid source satisfying the strong energy condition, the deformed Einstein equations yield a hairy Schwarzschild black hole solution
\begin{equation}\label{eq-static}
ds^2=-f(r)dt^2+\frac{dr^2}{f(r)}+r^2(d\theta^2+\sin^2\theta d\phi^2)
~~\mathrm{with}~~ f(r)=1-\frac{2M}{r}+\alpha e^{-r/(M-l_o/2)}.
\end{equation}
This metric describes a deformation of the Schwarzschild solution due to the presence of additional material sources, such as scalar hair, tensor hair, or fluidlike dark matter. When $\alpha=0$,  the metric reduces to the standard Schwarzchild solution in GR, corresponding to the absence of the matters.
Furthermore, in this scenario, hairy Kerr black hole was also derived, which is written in Boyer-Lindquist coordinates as
\begin{equation}
\begin{aligned}
ds^2 &= g_{tt}dt^2 + g_{rr}dr^2 + g_{\theta\theta}d\theta^2 + g_{\phi\phi}d\phi^2 + 2g_{t\phi}dtd\phi \\
&= -\left( \frac{\Delta - a^2 \sin^2 \theta}{\Sigma} \right) dt^2+\sin^2\theta\left(\Sigma+a^2 \sin^2\theta\left(2-\frac{\Delta-a^2\sin^2\theta}{\Sigma}\right)\right)d\phi^2\\
&+\frac{\Sigma}{\Delta}dr^2+\Sigma d\theta^2-2a\sin^2\theta\left(1-\frac{\Delta-a^2\sin^2\theta}{\Sigma}\right)dtd\phi,
\end{aligned}
\label{eq-metric}
\end{equation}
with
\begin{equation}
\begin{aligned}
\Sigma=r^2+a^2\cos^2\theta,~~~~~~\Delta=r^2+a^2-2Mr+\alpha r^2 e^{-r/\left(M-\frac{l_o}{2}\right)},
\end{aligned}
\end{equation}
describing a deformation of the Kerr black hole. Here, $M$ is the black hole mass, $\alpha$ is the deformation parameter and $l_o=\alpha l$ with $l$ being a length-scale parameter associated with the primary hair charge. The condition $l_o\leq 2M$ ensures asymptotic flatness.
Further details of these calculations can be found in \cite{Ovalle:2020kpd, Contreras:2021yxe}. For clarity, we have omitted the intermediate steps and directly presented the final hairy Schwarzschild and Kerr black hole metric. Moreover, in the following study, we will rescale all the physical quantities by the mass $M$ and set $M=1$ for evaluation.

\begin{figure}[H]
\centering
{\includegraphics[width=6.5cm]{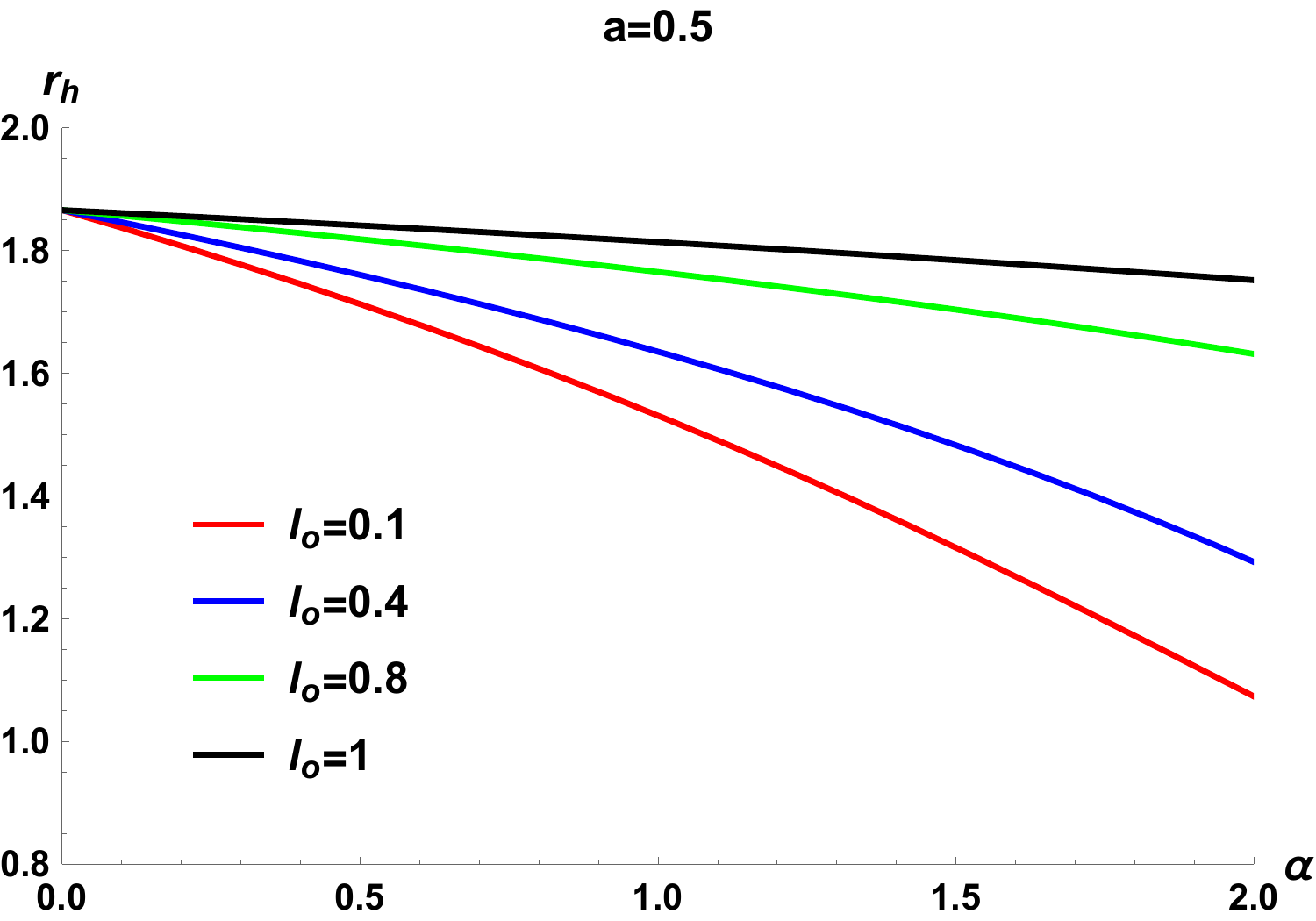}} \hspace{1cm}
{\includegraphics[width=6.5cm]{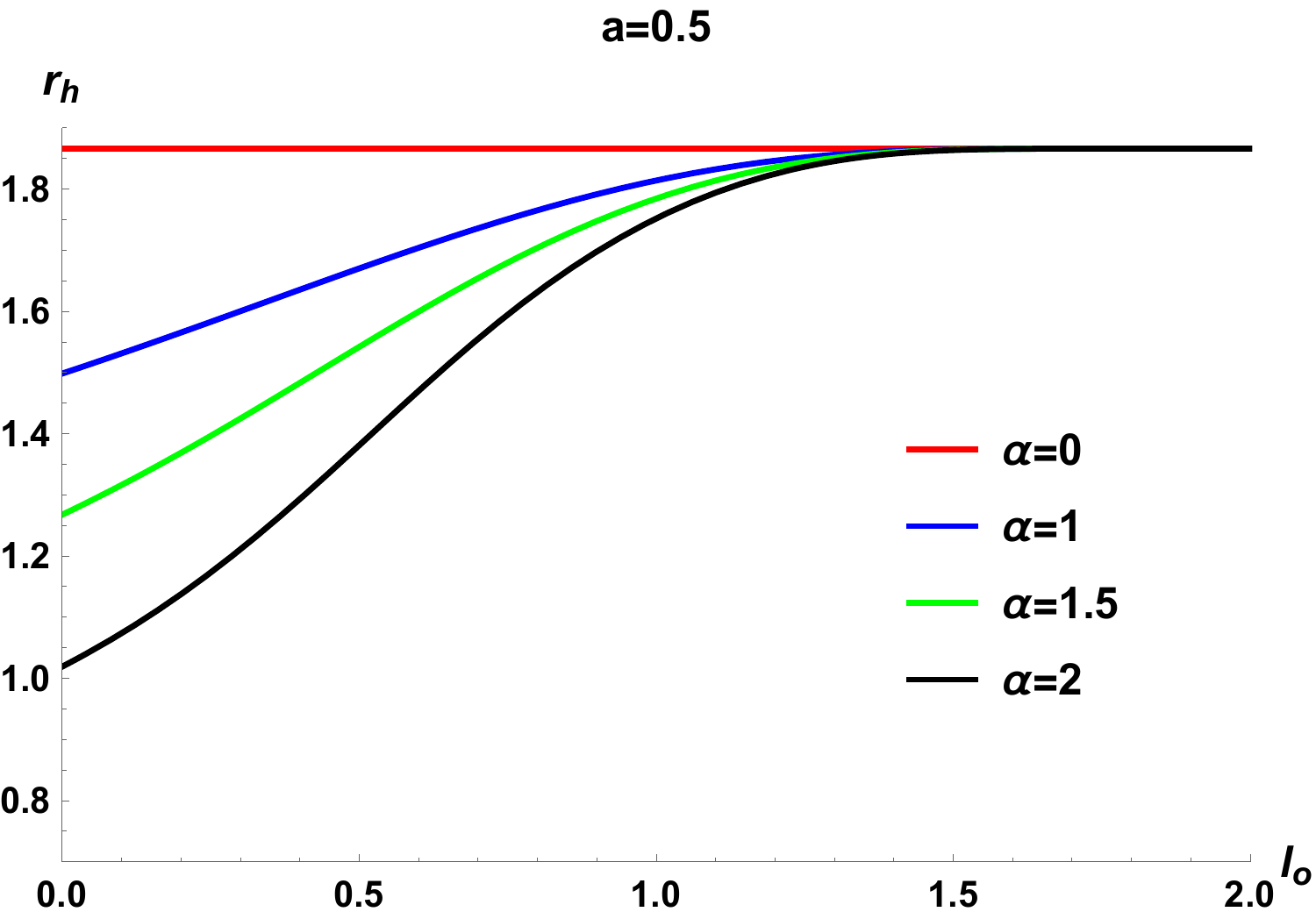}}
\caption{The event horizon $r_h$ of the hairy Kerr black hole as functions of the parameters $\alpha$ and $l_o$, respectively.}
\label{rh-a-lo}
\end{figure}

By solving $\Delta=0$, we can obtain the event horizon of the black hole. However, the expression of the event horizon is rather complicated and cannot be solved analytically. Instead, the influence of parameters $\alpha$ and $l_o$ on the event horizon is shown in FIG. \ref{rh-a-lo}. Obviously, the parameter $\alpha$  suppresses the black hole event horizon $r_h$, and as the parameter $l_o$ increases, the event horizon of the hairy Kerr black hole becomes larger. 

\subsection{Surrounding thin accretion disk }
We shall show a simple thin, Keplerian accretion disk consisting of free electrically neutral plasma, of which the motion is solely controlled by gravitational field of the black hole. This means that the motion of the accretion material admits timelike geodesic equations around the hairy Kerr black hole, which can be derived from the Lagrangian of massive particle
\begin{eqnarray}\label{eq-lagrangian}
\mathcal{L} = \frac{1}{2} \left( g_{tt} \dot{t}^2 + 2 g_{t\phi} \dot{t} \dot{\phi}
+ g_{rr} \dot{r}^2 + g_{\theta\theta} \dot{\theta}^2 + g_{\phi\phi} \dot{\phi}^2 \right) \, ,
\end{eqnarray}
where $\tau$ is the proper time of the accreting material and the dot denotes the derivative with respect to $\tau$. From the Lagrangian, we can derive two constants of the motion
\begin{eqnarray}\label{eq-pt}
\frac{\partial \mathcal{L}}{\partial \dot{t}} = g_{tt} \dot{t} + g_{t\phi} \dot{\phi} = - E \, , \\
\frac{\partial \mathcal{L}}{\partial \dot{\phi}} = g_{t\phi} \dot{t} + g_{\phi\phi} \dot{\phi} = L \, . \label{eq-pphi}
\end{eqnarray}
where $E$ and $L$ are the conserved  energy  and  axial component of the angular momentum for per unit mass.
Subsequently, we can obtain the {equations} of $t$ and $\phi$ as
\begin{eqnarray}\label{eq-tdot}
\dot{t} &=& \frac{g_{\phi\phi} E + g_{t\phi} L}{g_{t\phi}^2 - g_{tt} g_{\phi\phi}} \, , \\
\dot{\phi} &=& - \frac{g_{t\phi} E + g_{tt} L}{g_{t\phi}^2 - g_{tt} g_{\phi\phi}} \, .
\label{eq-phidot}
\end{eqnarray}
It is noted that the above equations should be satisfied by arbitrary timelike orbits of massive particles in the spacetime. To describe the motion of the component of an accretion disk, we shall consider that it is infinitesimally thin and perpendicular to the black hole spin axis. In the following studies, we will set $\theta = \pi/2$ and $\dot{\theta} = 0$.

\subsubsection{$r\geq r_{\rm ISCO}$}

Since the components of the accretion disk follow timelike geodesics and 
their motions are determined by the gravitational field of the black hole, the inner edge of the accretion disk can naturally be at the radius of the innermost stable circular orbit (ISCO)~\cite{Bardeen:1972fi}. 
Outside the ISCO, the motions of the components in the disk can be approximated with equatorial circular orbits which further have
\begin{eqnarray}\label{eq-rdd=0}
\dot{r}=\ddot{r}=0.
\end{eqnarray}
Then the geodesic equations of the massive particles are written as 
\begin{eqnarray}
\left( \partial_r g_{tt} \right) \dot{t}^2 + 2 \left( \partial_r g_{t\phi} \right)
\dot{t} \dot{\phi} + \left( \partial_r g_{\phi\phi} \right) \dot{\phi}^2 = 0 \, .
\end{eqnarray}
The angular velocity of the disk component measured in the Boyer-Linguist coordinate is 
\begin{eqnarray}\label{eq-omega}
\Omega_{\rm K} = \frac{d\phi}{dt}  = \frac{ - \partial_r g_{t\phi} \pm \sqrt{ \left( \partial_r g_{t\phi} \right)^2
- \left( \partial_r g_{tt} \right) \left( \partial_r g_{\phi\phi} \right) }}{\partial_r g_{\phi\phi}} \, ,
\end{eqnarray}
where the sign “+” is for co-rotating orbits and the sign “-” is for counter-rotating orbits.

\begin{figure}[H]
\centering
\subfigure[\, ]
{\includegraphics[width=6.5cm]{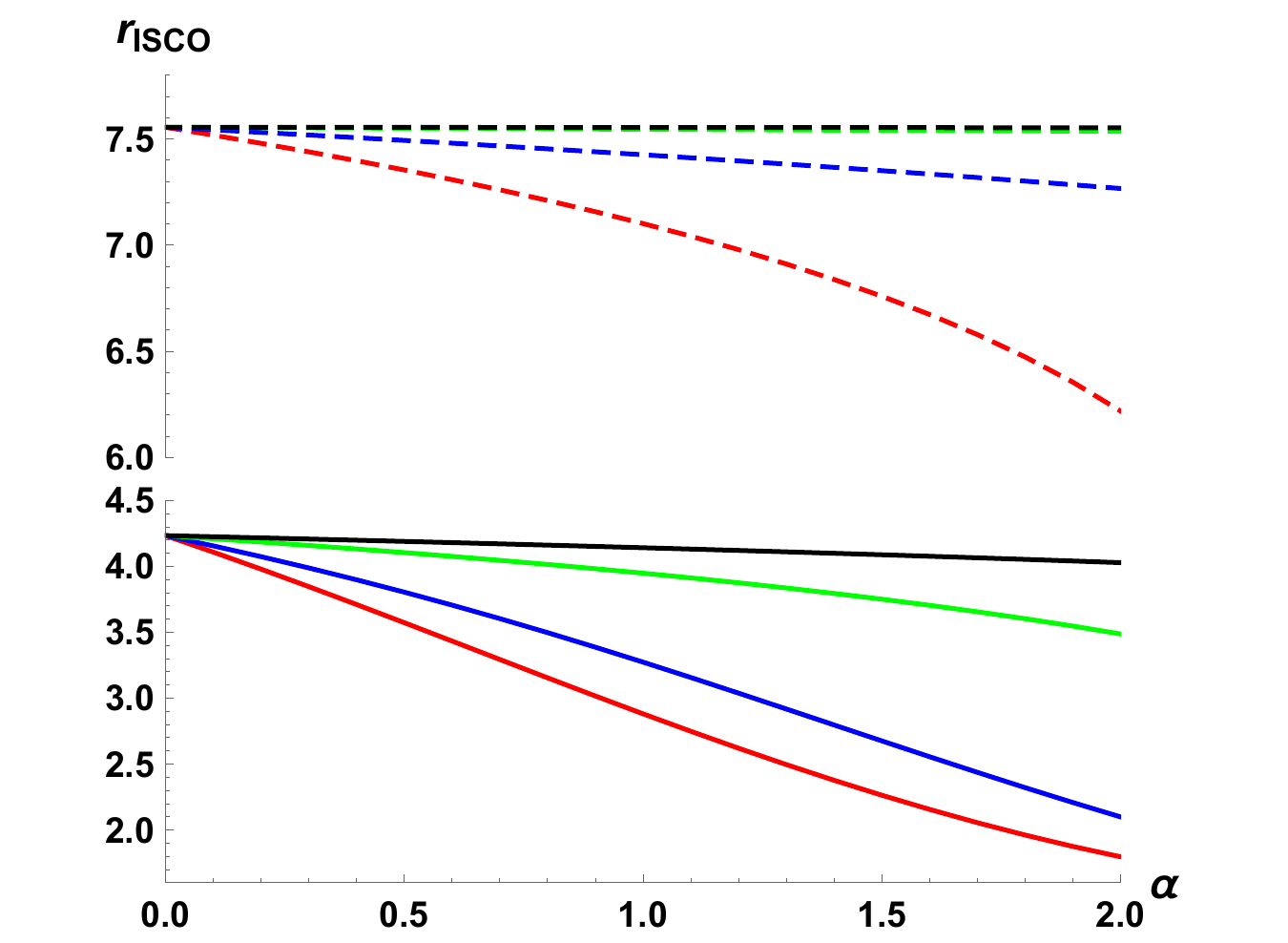}} \hspace{5mm}
\subfigure[\, ]
{\includegraphics[width=6cm]{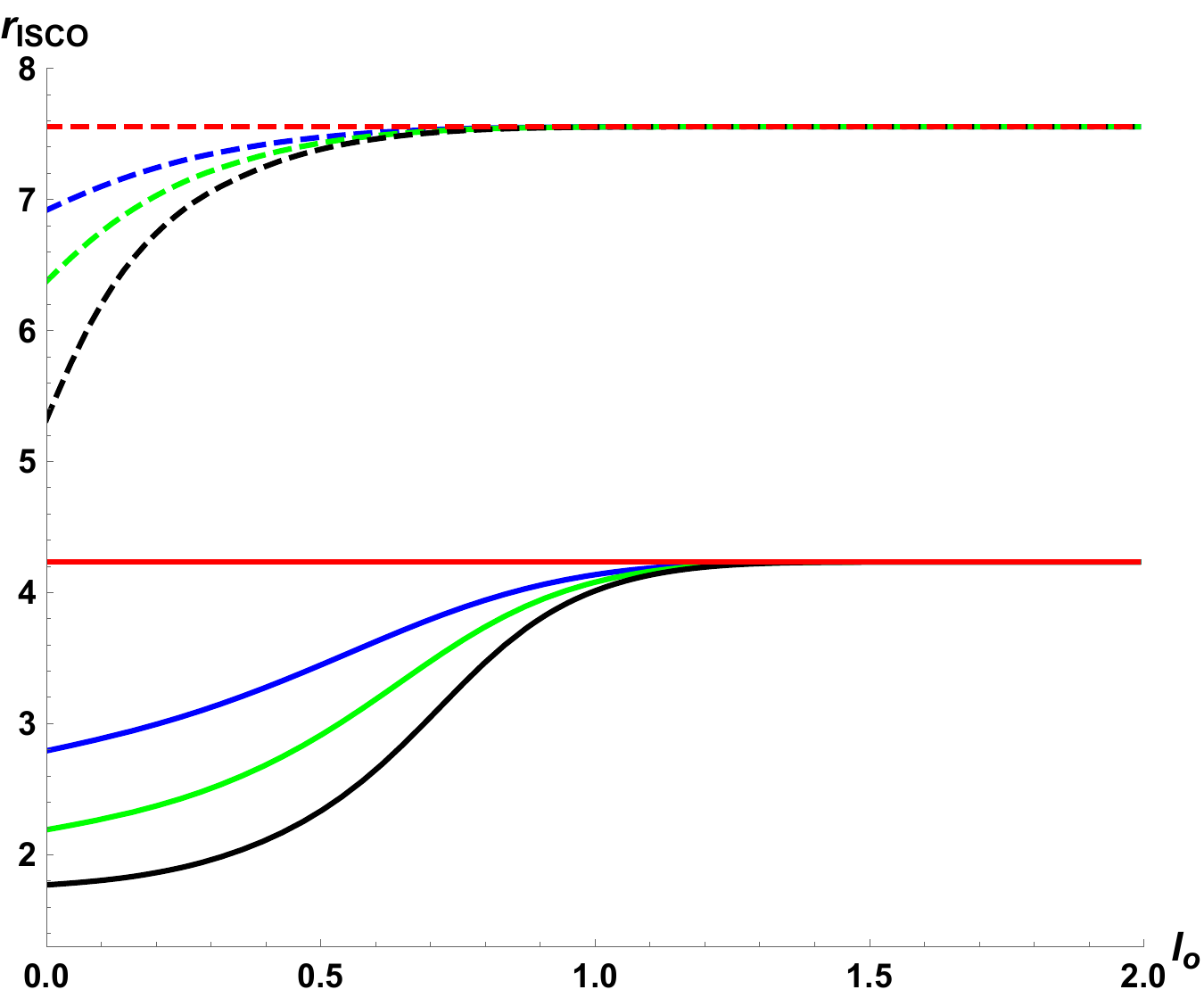}} \\
\subfigure[\, ]
{\includegraphics[width=6.5cm]{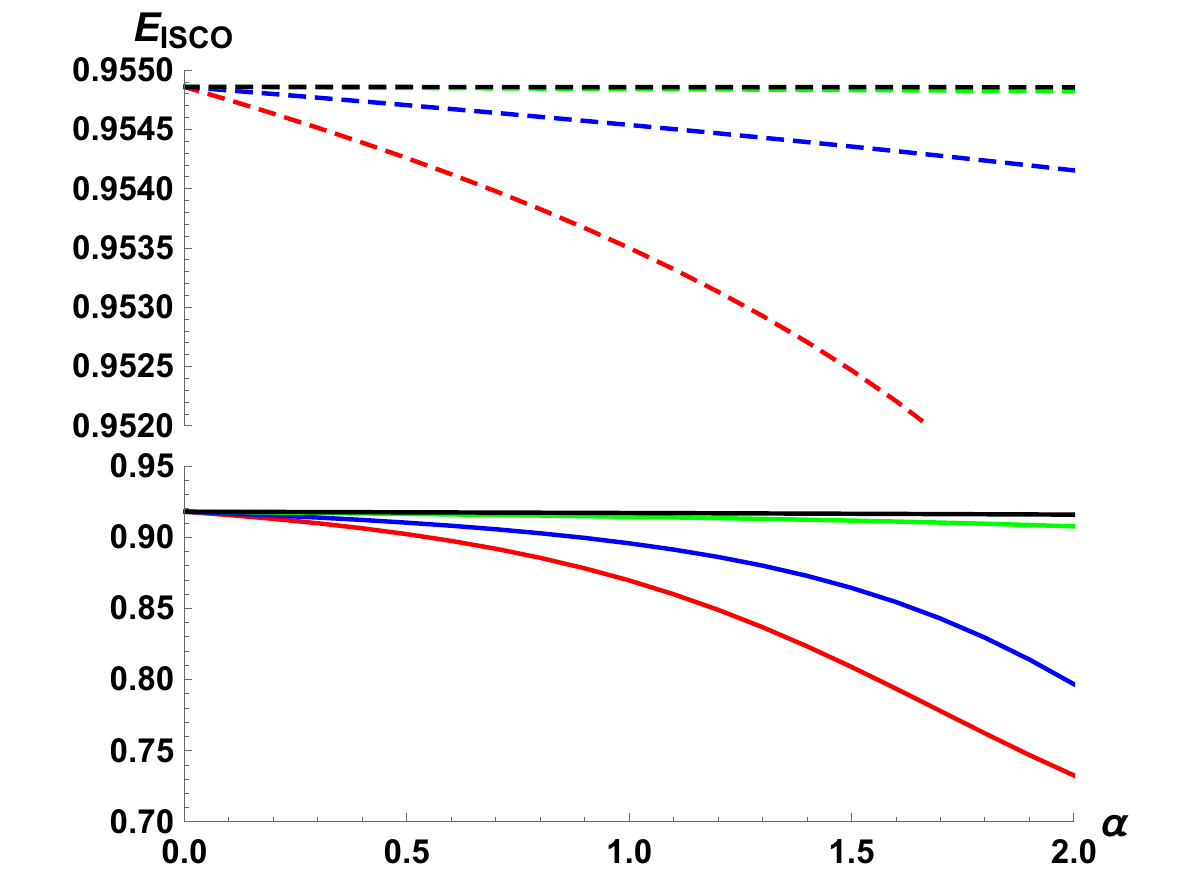}} \hspace{5mm}
\subfigure[\, ]
{\includegraphics[width=6.5cm]{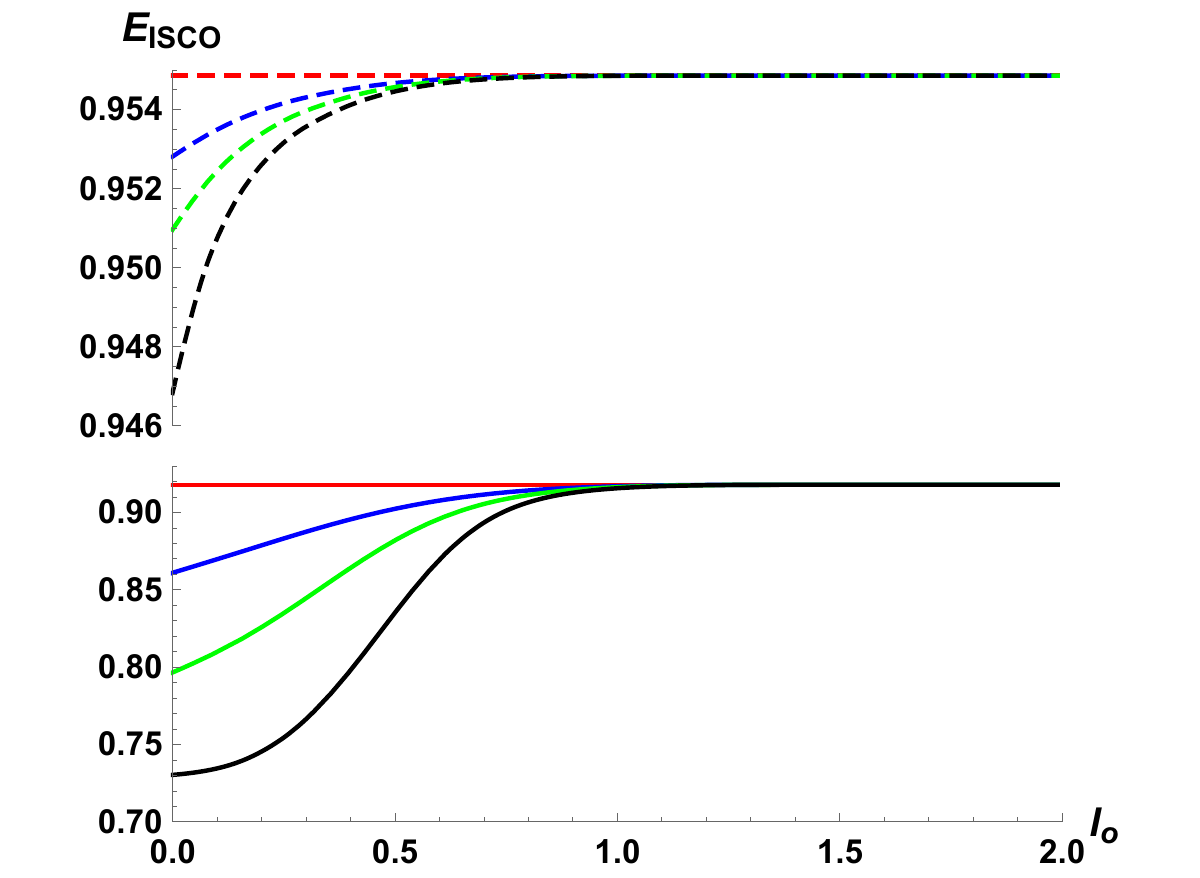}}\\
\subfigure[\, ]
{\includegraphics[width=6.5cm]{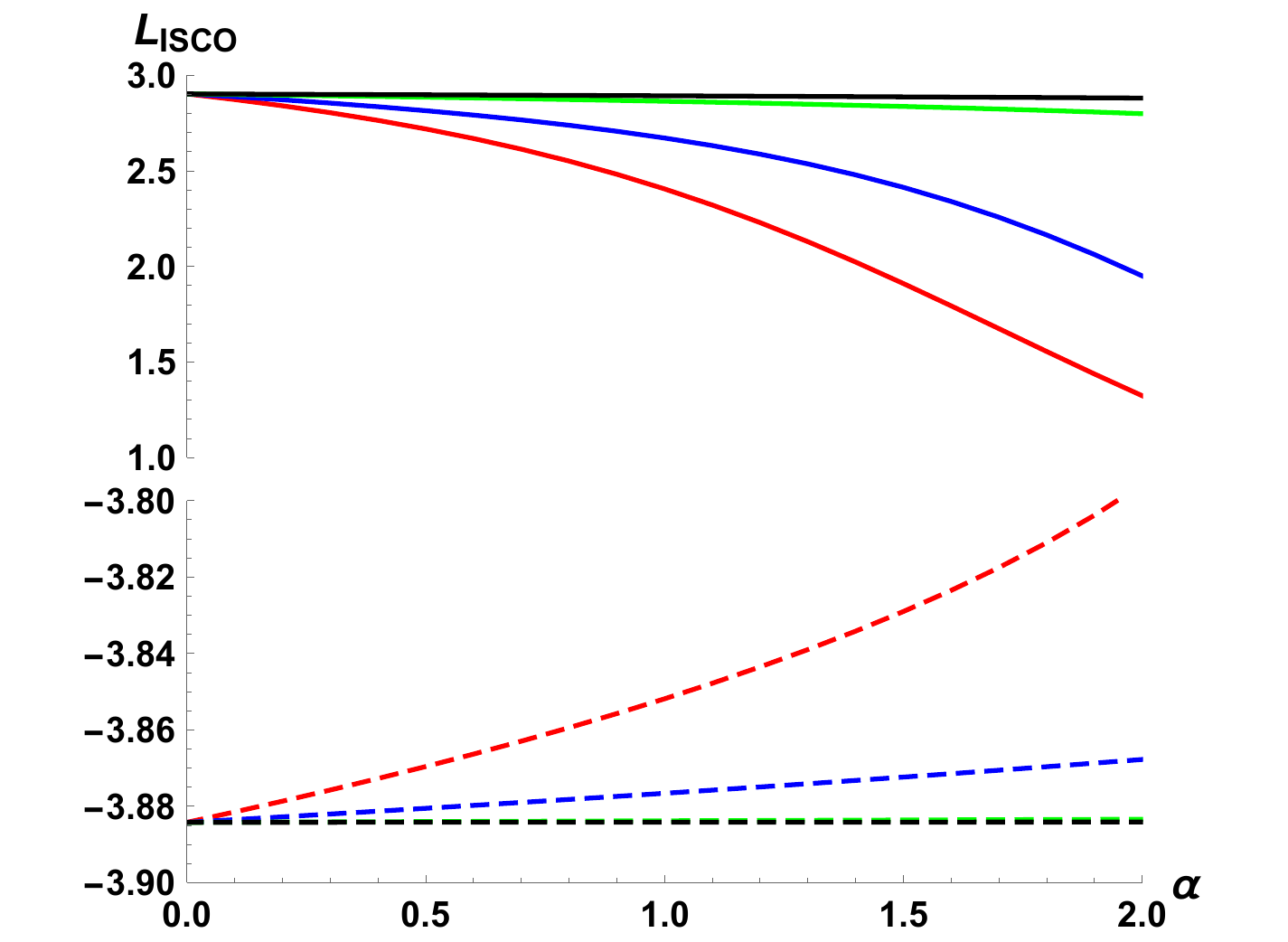}} \hspace{5mm}
\subfigure[\, ]
{\includegraphics[width=6.5cm]{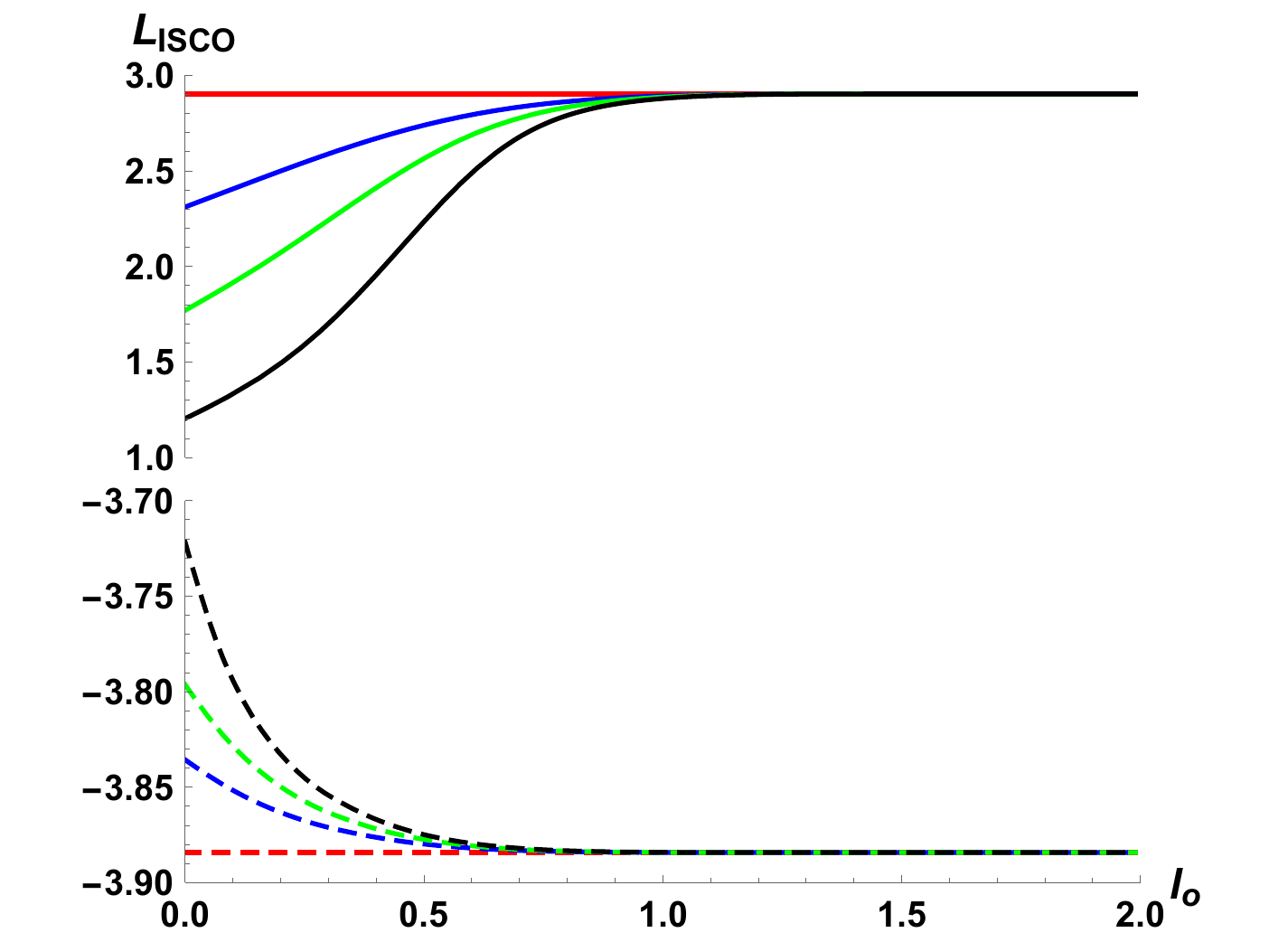}}
\caption{The $r_{\rm ISCO}$, $E_{\rm ISCO}$  and $L_{\rm ISCO}$ of the massive particle orbiting around the hairy Kerr black hole with $a=0.5$ as functions of the parameters $\alpha$ (left column) and $l_o$ (right column). In each plot, the solid and dashed curves are for the prograde and retrograde orbits, respectively. {\bf Left}: the red, blue, green and black curves correspond to $l_o=0.1$, $0.4$, $0.8$ and $1$, respectively. {\bf Right}: the red, blue, green and black curves correspond to $\alpha=0$, $1$, $1.5$ and $2$, respectively.}
\label{risco-a-lo}
\end{figure}

We proceed to find the ISCO radius by studying the stability of the geodesic circular orbits. Combining $g_{\mu\nu} \dot{x}^\mu \dot{x}^\nu = -1$ for timelike particle, \eqref{eq-tdot}, \eqref{eq-phidot} and $\theta=\pi/2$, we can obtain the radial motion equation
\begin{eqnarray}
g_{rr} \dot{r}^2 - V_{\rm eff} (r)=0 \, ,
\end{eqnarray}
with
\begin{eqnarray}
V_{\rm eff}(r) = \frac{g_{\phi\phi} E^2 + 2 g_{t\phi} E L_z + g_{tt} L^2}{g_{t\phi}^2 - g_{tt} g_{\phi\phi}} - 1 \, .
\end{eqnarray}
Then the conditions \eqref{eq-rdd=0} for the circular orbits are converted  to 
$V_{\rm eff}=V_{\rm eff}'=0$ where the prime denotes the derivative with respect to $r$,
but the (in)stability of the circular orbit is determined by the sign of $V_{\rm eff}''$, i.e,  {$V_{\rm eff}''> 0$ } corresponds to stability while  $V_{\rm eff}''<0$ gives unstable circular orbit. 
Then the ISCO radius separating stable orbits ($r > r_{\rm ISCO}$) and unstable orbits ($r < r_{\rm ISCO}$) satisfies 
\begin{eqnarray}
V_{\rm eff}=V_{\rm eff}'=V_{\rm eff}''=0,
\end{eqnarray}
from which we can solve out $r_{\rm ISCO}$, $E_{\rm ISCO}$ and $L_{\rm ISCO}$. 
{For the hairy Kerr black hole, it is difficult to give analytical expressions for those quantities of the ISCO, so we depict our results of the prograde and
retrograde particles in FIG. \ref{risco-a-lo} for different parameters $l_o$ and $\alpha$. Obviously,  the first row shows that the parameter $\alpha$ suppresses the radius of the ISCO,  similar to the event horizon of hairy Kerr black hole. While as the parameter $l_o$ increases, the ISCO radius of the black hole increase. In addition, the ISCO for the retrograde particle is larger than that for the prograde particle as expected. The behavior of  $E_{\rm ISCO}$ is similar to $r_{\rm ISCO}$, as shown in the second row. $L_{\rm ISCO}$ depicted in third row seems to behave differently from  $r_{\rm ISCO}$ and  $E_{\rm ISCO}$, but they indeed have similar behaviors if we take the absolute value of  $L_{\rm ISCO}$ for retrograde orbits.}

\subsubsection{$r_h<r< r_{\rm ISCO}$}

Inside the ISCO, the circular orbits are unstable, so the component can quickly plunge into the black hole. So in this region $r_{h} < r < r_{\rm ISCO}$, we can consider that the disk components carry out plunging motion with the conserved  energy $E=E_{\rm ISCO}$ and angular momentum $L=L_{\rm ISCO}$. Therefore, the motions of the disk component are determined by
\begin{eqnarray}\label{eq-tdot-plunging}
\dot{t} &=& \frac{g_{\phi\phi} E_{\rm ISCO} + g_{t\phi} L_{\rm ISCO}}{g_{t\phi}^2 - g_{tt} g_{\phi\phi}} \, , \\
\dot{\phi} &=& - \frac{g_{t\phi} E_{\rm ISCO} + g_{tt} L_{\rm ISCO}}{g_{t\phi}^2 - g_{tt} g_{\phi\phi}} \, ,
\label{eq-phidot-plunging}
\end{eqnarray}
and 
\begin{equation}
\begin{aligned}
\dot{r}=-\sqrt{\frac{V_{\rm eff}}{g_{rr}}}\Bigg|_{E=E_{\rm ISCO},\, L= L_{\rm ISCO}}\label{orbits},
\end{aligned}
\end{equation}
where we choose the minus sign because the component of the disk is falling into the black hole.

\section{Images  of hairy Kerr black hole illuminated by a thin accretion disk}\label{sec-image}

In this section we shall figure out the optical appearance of the accretion disk described in previous section. To this end, we will firstly review the method of taking  photographs by  a distant observer employed in \cite{Cunningham:1975zz,Gralla:2020srx,Hou:2022eev}, and then analyze the effect of the hair on the optical appearance of the rotating black hole surrounded by the accretion disk.

\subsection{Review on the setup and method}
We proceed to study  the photons orbiting  from the accretion disk surrounding the hairy Kerr black hole to a distant observer. It is convenient to consider a zero-angular-momentum observer (ZAMO) positioned at $(t_o=0,r_o,\theta_o,\phi_o=0)$ due to the symmetries in the $t$ and $\phi$ directions of the spacetime. To describe this observer’s local frame, we consider an orthonormal tetrad \cite{Hou:2022eev}
\begin{equation}
\begin{aligned}
& e_{(0)} = \sqrt{\frac{-g_{\phi\phi}}{g_{tt}g_{\phi\phi}-g^2_{t\phi}}}\left(1, 0, 0, -\frac{g_{t\phi}}{g_{\phi\phi}}\right), \quad e_{(1)} = \left( 0, -\frac{1}{\sqrt{g_{rr}}}, 0, 0 \right), \\
& e_{(2)} = \left( 0, 0, \frac{1}{\sqrt{g_{\theta\theta}}}, 0 \right), \quad e_{(3)} = \left( 0, 0, 0, -\frac{1}{\sqrt{g_{\phi\phi}}} \right).
\end{aligned}
\label{eq-ZAMO}
\end{equation}
Note that the  minus sign in $e_{(1)}$ and $e_{(3)}$ allows us to facilitate the backward ray-tracing method. In order to analyze the light rays near the hairy Kerr black hole, from the Lagrangian of the photons
\begin{equation}
\begin{aligned}
\mathcal{L}=\frac{1}{2}g_{\mu\nu}\dot{x}^\mu\dot{x}^\nu
=\frac{1}{2}(g_{tt}\dot{t}^2+g_{rr}\dot{r}^2+g_{\theta\theta}\dot{\theta}^2+g_{\phi\phi}\dot{\phi}^2+2g_{t\phi}\dot{t}\dot{\phi}),
\end{aligned}
\end{equation}
we can derive the four-momentum  $k_\nu=\partial \mathcal{L}/\partial \dot{x}^\nu$,
\begin{equation} \label{eq-k_mu}
\begin{aligned}
k_t=g_{t\phi}\dot{\phi}+g_{tt}\dot{t}=-\varepsilon, ~~~~k_r=g_{rr} \dot{r},~~~k_\theta=g_{\theta\theta}\dot{\theta}, ~~~k_\phi=g_{\phi\phi}\dot{\phi}+g_{t\phi}\dot{t}=\mathbb{L},
\end{aligned}
\end{equation}
where the dot represent derivative with respect to the affine parameter $\lambda$ of the null geodesic, and  $\varepsilon$ and $\mathbb{L}$ are the conserved energy and axial component of angular momentum of the photons. Subsequently, in the ZAMO frame, the four-momentum of the photons is given by
\begin{equation}\label{eq-p_mu}
\begin{aligned}
p_{(\mu)}=k_\nu e^\nu_{(\mu)}~~~~~\mathrm{and} ~~~~~p^{(\mu)}=\eta^{\mu\sigma}e_{(\sigma)}^{\nu} k_\nu ,
\end{aligned}
\end{equation}
where $e_{(\mu)}^{\nu} $ is defined in \eqref{eq-ZAMO}. Further defining celestial coordinates $\Theta$ and $\Psi$ in ZAMO frame to label the light ray,  and considering that the tangent vector of light ray can be measured  both in the frame of the observer and in the Boyer-Lindquist coordinates \cite{Grenzebach:2014fha,Hu:2020usx}, we have the following relation
\begin{equation}
\begin{aligned}
\cos \Theta=\frac{p^{(1)}}{p^{(0)}},~~~~ \tan\Psi=\frac{p^{(3)}}{p^{(2)}}.
\end{aligned}
\end{equation}
Using the stereographic projection, we can then display the image on a two-dimensional screen of the ZAMO by  projecting  the celestial coordinates into a Cartesian coordinates via
\begin{equation}
\begin{aligned}
x=-2\tan\frac{\Theta}{2}\sin\Psi,~~~~y=-2\tan\frac{\Theta}{2}\cos\Psi.
\end{aligned}
\end{equation}
Thus, for photon with given initial position $(0,r_o,\theta_o,0)$ and initial momentum, we can 
integrate the null geodesic equation to trace  the trajectory  of the light ray  backward to infinity or the outer horizon.

When the light is traced back from the observer, it may cross the accretion disk on the equatorial plane many times. Each time the light ray intersects with the accretion disk, it gains additional energy and contributes to the observed intensity.  So it is important to determine  the radius of its intersection point $r_n(x,y)$ in the tracing procession, where  $n=1,2,3....$ is the number of intersections.  It is noted  that  $r_n(x,y)$ is known as the transfer function, which directly determines the image shape of the accretion disk with $n-$th intersections. Concretely, $n=1$ gives the direct emission image, $n=2$ gives the lensed ring emission image, and $n\geq3$ gives the photon ring emission image  \cite{Gralla:2019xty}.

Next, in order to figure out the image of the accretion disk on the screen in the frame of the ZAMO, we shall study the observed intensity, which would be different from the emission intensity because of redshift effect and, emission and absorption during the intersections between the light rays and disk. To simplify the analysis, we consider an accretion disk model with the change in the intensity \cite{Lindquist:1966igj}.
\begin{equation}
\begin{aligned}
\frac{d}{d\lambda}\left(\frac{I_v}{v^3}\right)=\frac{J_v-\kappa_v I_v}{v^2},
\label{intensity}
\end{aligned}
\end{equation}
where  $I_{v}$, $J_{v}$, and $\kappa_{v}$ correspond to the specific intensity, emissivity, and absorption coefficient at the frequency $v$, respectively.  We consider that the accretion disk is steady, axisymmetic, and has $Z_2$ symmetry about the
equatorial plane, and it is optically and geometrically thin such that the absorption can be ignored.  Therefore,  we can integrate \eqref{intensity} along the
traced-back trajectories and reduce the intensity at each location on the observer’s
screen as \cite{Chael:2021rjo,Hadar:2020fda}
\begin{equation}
\begin{aligned}
I_{\nu o}=\sum_{n=1}^{N_{\text{max}}} f_n g^3(r_n) J_{e}(r_n),
\end{aligned}
\label{intensity-z}
\end{equation}
where the  redshift factor is $g(r_n)=v_o/v_e$, $f_n$ is  the ‘fudge factor’ that controls the brightness of the higher-order photon ring  and $J_{e}(r_n)$ is the emissivity function per unit volume at a given frequency. We normalize
all the fudge factors to be $1$ as done in \cite{Hou:2022eev}.  Since the images of M87* and Sgr A* were captured at an observing wavelength of 1.3 mm (230 GHz), we follow \cite{Chael:2021rjo} to choose  the emissivity function as
\begin{equation}
\begin{aligned}
J_e(r)=\exp\left[-\frac{1}{2}\left(\log\frac{r}{r_h}\right)^2-2\left(\log\frac{r}{r_h}\right)\right].
\end{aligned}
\end{equation}
The redshift factor $g(r_n)$ for our accretion disk that follows circular orbits down to the ISCO, beyond which the  matter  infall along geodesics with conserved quantities equal to ISCO, was carried out in \cite{Cunningham:1975zz}. Its derivation  was later reviewed in \cite{Gralla:2020srx,Hou:2022eev}, and they are calculated  as
\begin{equation}
g(r_n)=\begin{cases}
\frac{p_{(0)}/k_t}{\zeta(1-( b\Omega_K)} \Big|_{r=r_n} \quad \text{for $r_n>r_{\rm ISCO}$}\\
-\frac{p_{(0)}/k_t}{\dot{r} k_r/\varepsilon+E_{\rm ISCO}(g^{tt}-g^{t\phi}b)+L_{\rm ISCO}(g^{\phi\phi}b-g^{t\phi})}\Big|_{r=r_n} \quad \text{for  $r_h<r_n<r_{\rm ISCO}$}
\end{cases}
\end{equation}
where $\dot{r}$  is radial velocity of disk component in \eqref{orbits}, and 
\begin{equation}
\zeta=\sqrt{\frac{-1}{g_{tt}+2g_{t\phi}\Omega_K
+g_{\phi\phi}\Omega^2_K}},
\end{equation}
with $\Omega_K$ evaluated in \eqref{eq-omega}, and 
\begin{equation}
b=\frac{\mathbb{L}}{\varepsilon}=\frac{k_\phi}{-k_t}
\end{equation}
represents the impact parameter of the photon with $k_{\mu}$ and $p_{(\mu)}$ defined in \eqref{eq-k_mu} and \eqref{eq-p_mu}, respectively.

\subsection{Effects of hair on the observational signatures}
With the strategy described in previous subsection, we are ready to numerically simulate the  images of hairy Kerr black hole illuminated by the mentioned accretion disk. Since the effects of the spinning parameter, rotating direction of accretion and  inclination angles of observer on the black hole images have been elaborately analyzed in \cite{Hou:2022eev,Wang:2023fge,He:2024amh,Zheng:2024brm,Li:2024ctu,Yang:2024nin,Wang:2024uda,Guo:2024mij}, here we shall mainly concentrate on the effects of black hole hair on the observational properties, such that  we fix the observation distance {$r_o=100$} and $\theta_{o}=80^{\circ}$, and extend  the  prograde  accretion disk from $r_{\rm ir}=r_h$ to the outer radius of {$r_{\rm or}=20$}.

\begin{figure}[H]
\centering
\subfigure[\, $\text{Kerr},a=0.1$]
{\includegraphics[width=5cm]{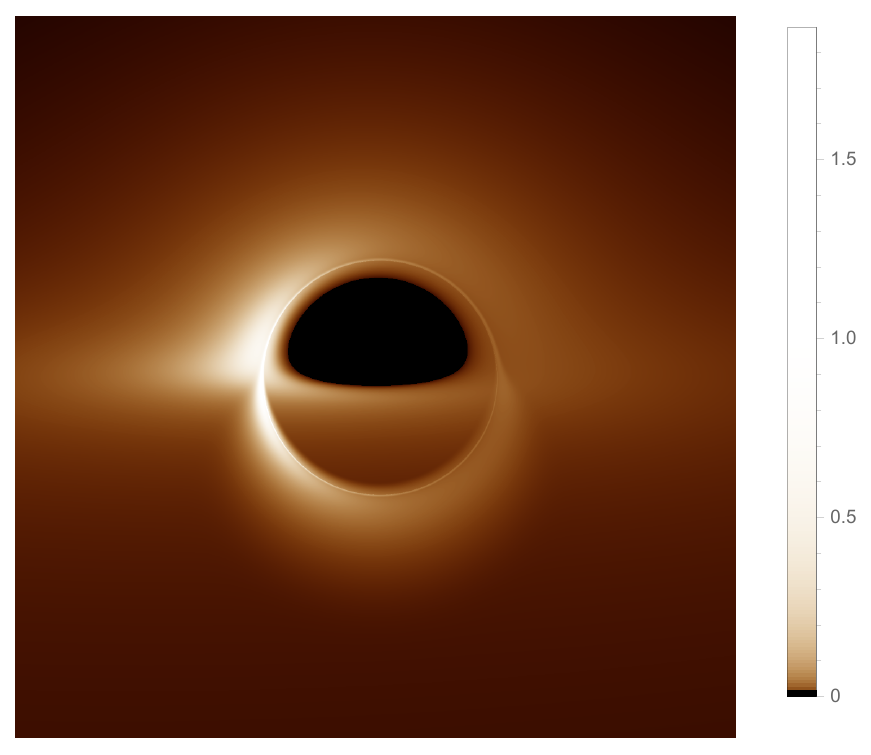}} \hspace{2mm}
\subfigure[\, $a=0.1,\alpha=4,l_o=0.8$]
{\includegraphics[width=5cm]{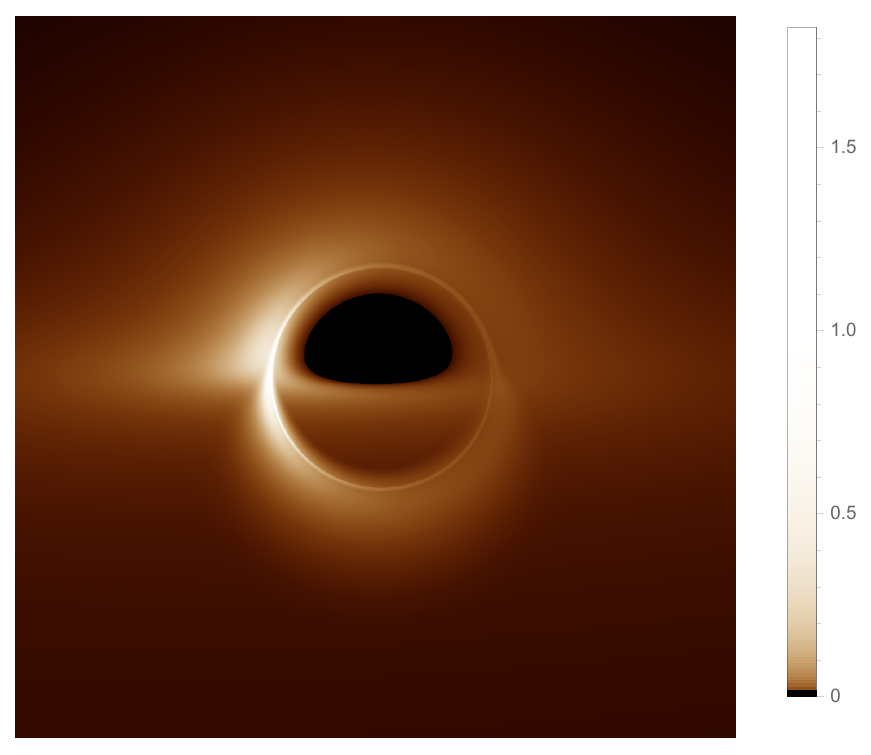}}\hspace{2mm}
\subfigure[\, $a=0.1,\alpha=4,l_o=0.5$]
{\includegraphics[width=5cm]{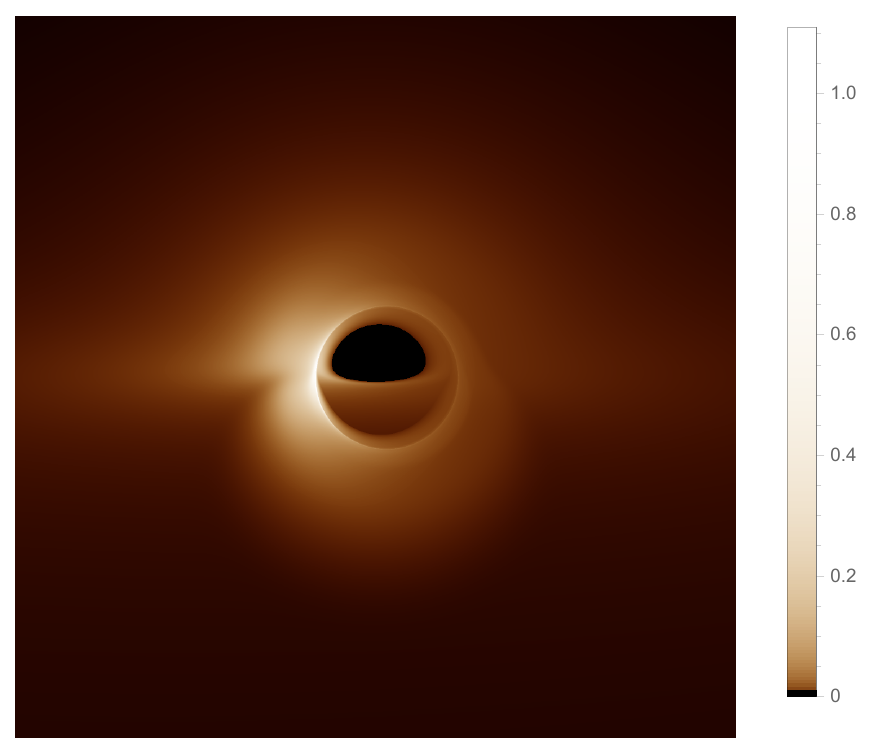}}\\
\subfigure[\, $\text{Kerr},a=0.5$]
{\includegraphics[width=5cm]{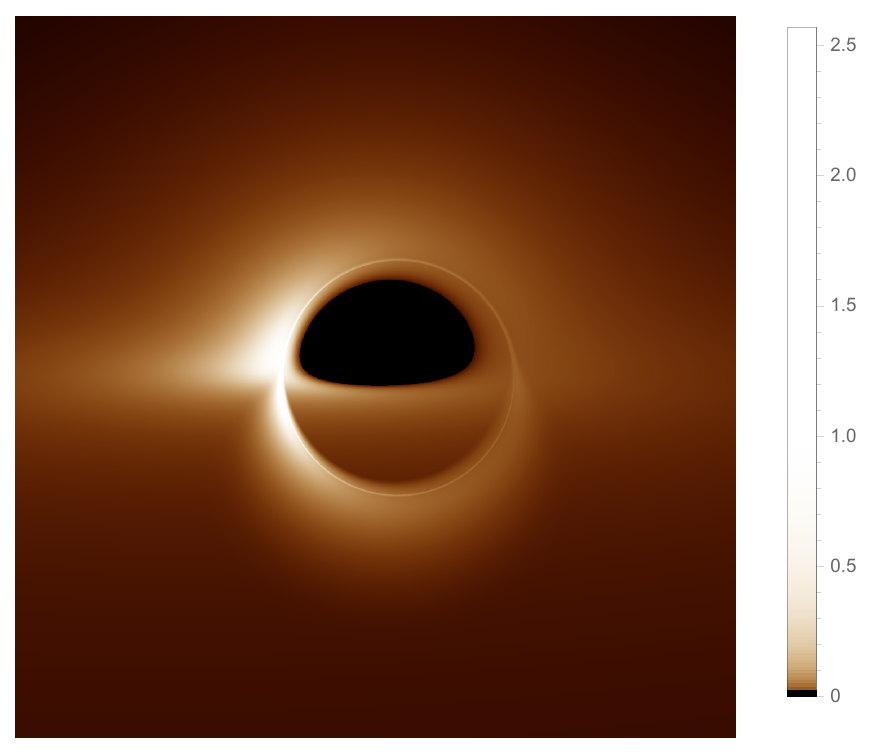}} \label{}\hspace{2mm}
\subfigure[\, $a=0.5,l_o=1,\alpha=3$]
{\includegraphics[width=5cm]{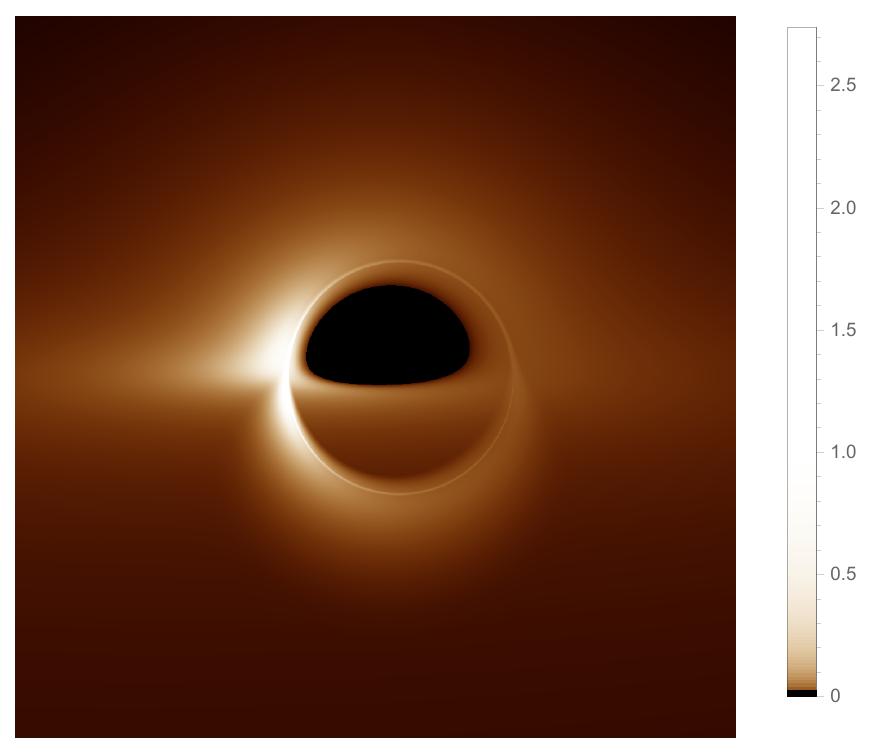}}\hspace{2mm}
\subfigure[\, $a=0.5,l_o=1,\alpha=5$]
{\includegraphics[width=5cm]{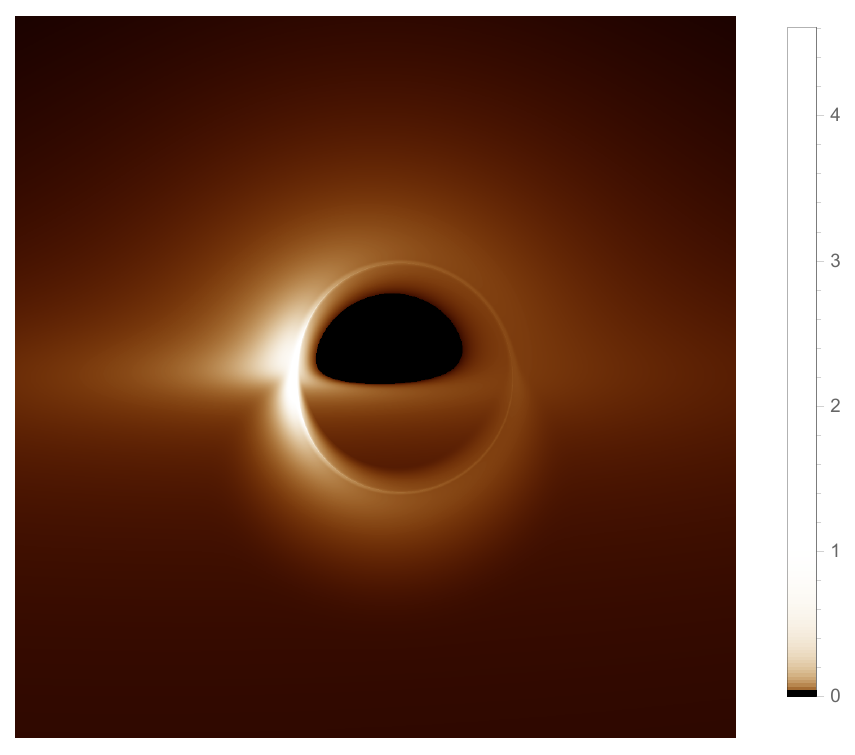}}\\
\caption{Images of the (hairy) Kerr black hole illuminated by prograde accretion flows for selected model parameters. }
\label{fig-Images}
\end{figure}

\begin{figure}[H]
\centering
\subfigure[\, $a=0.1$]
{\includegraphics[width=6cm]{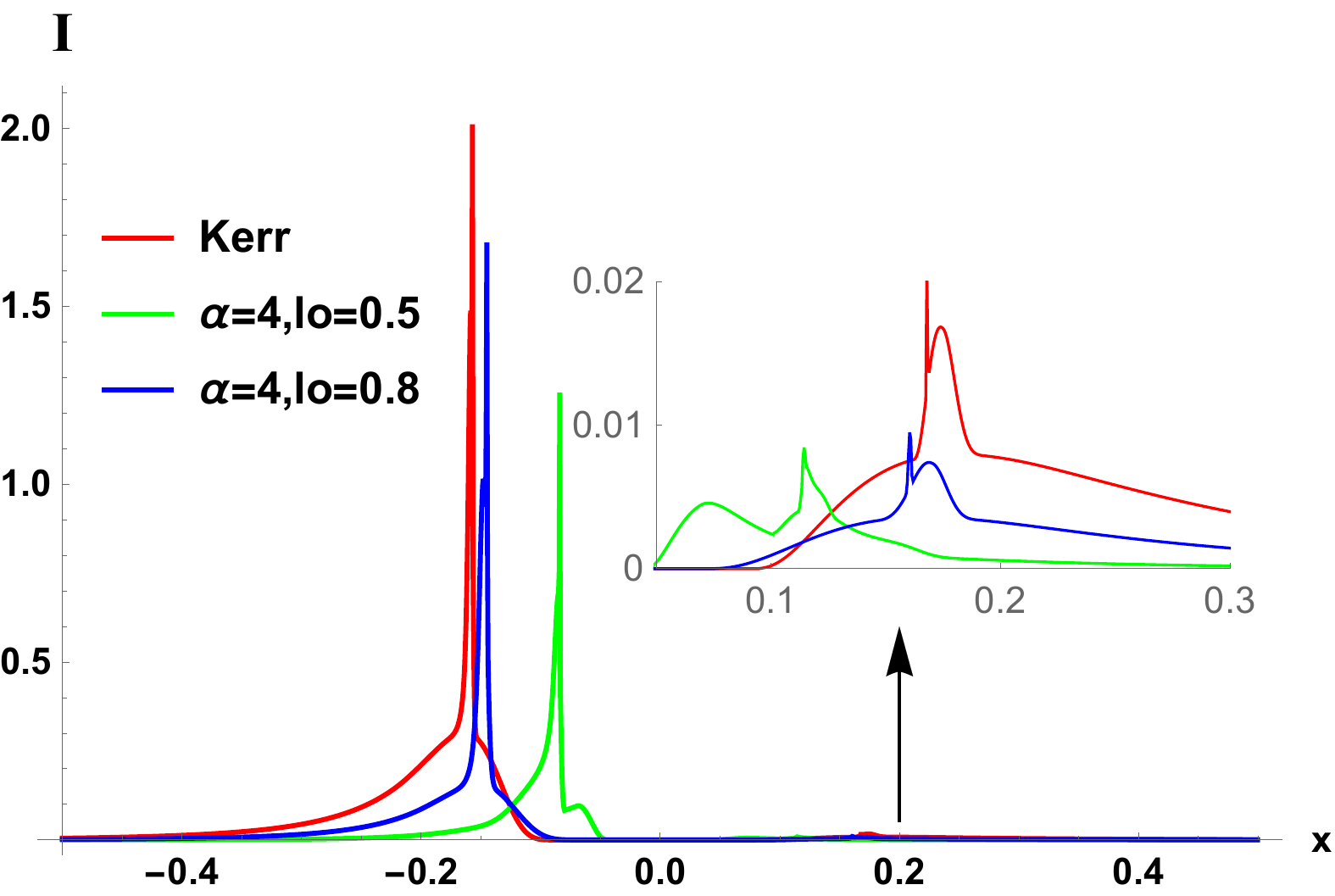}} \hspace{2mm}
\subfigure[\, $a=0.5$]
{\includegraphics[width=6cm]{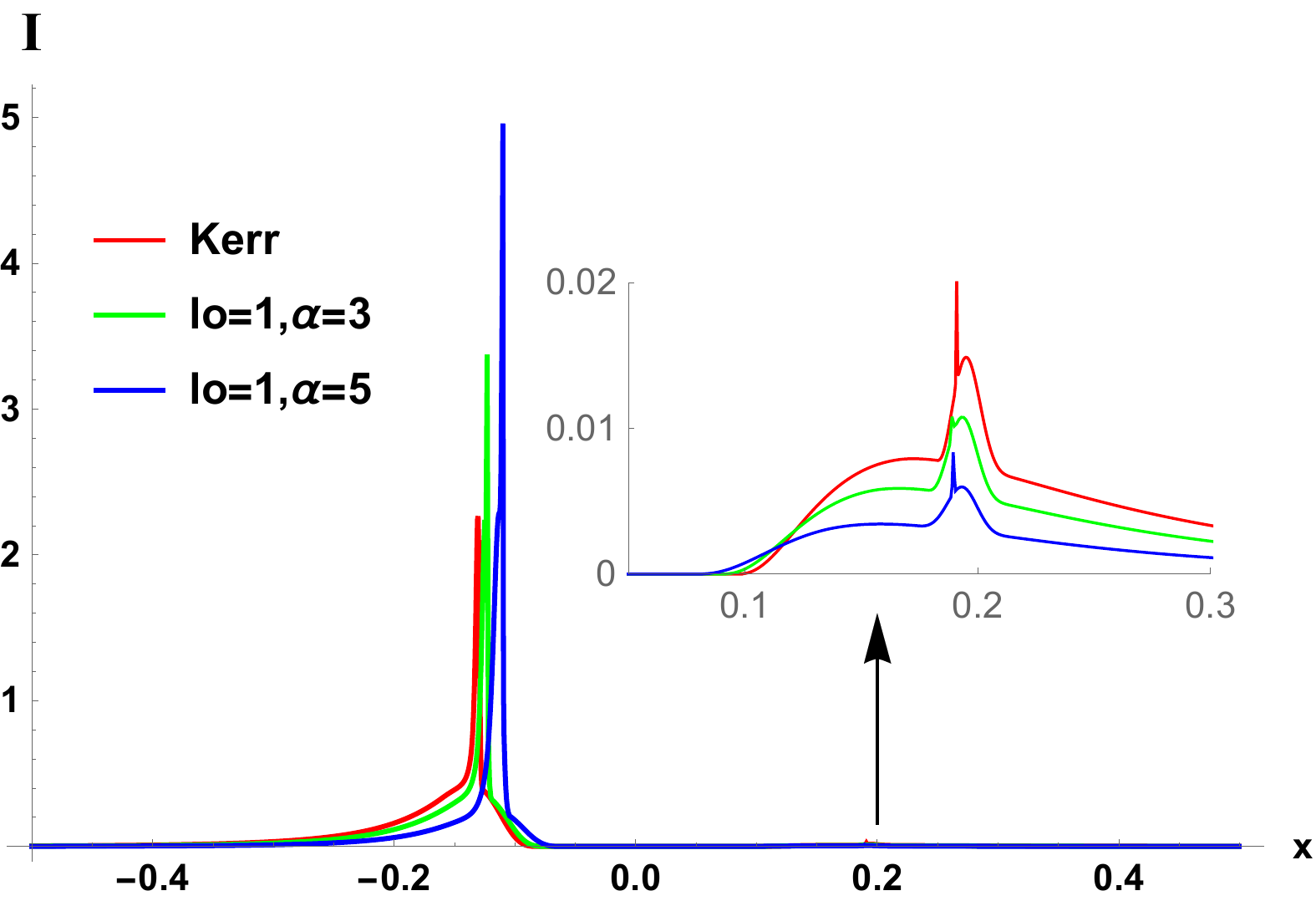}}\hspace{2mm}\\
\subfigure[\, $a=0.1$]
{\includegraphics[width=6cm]{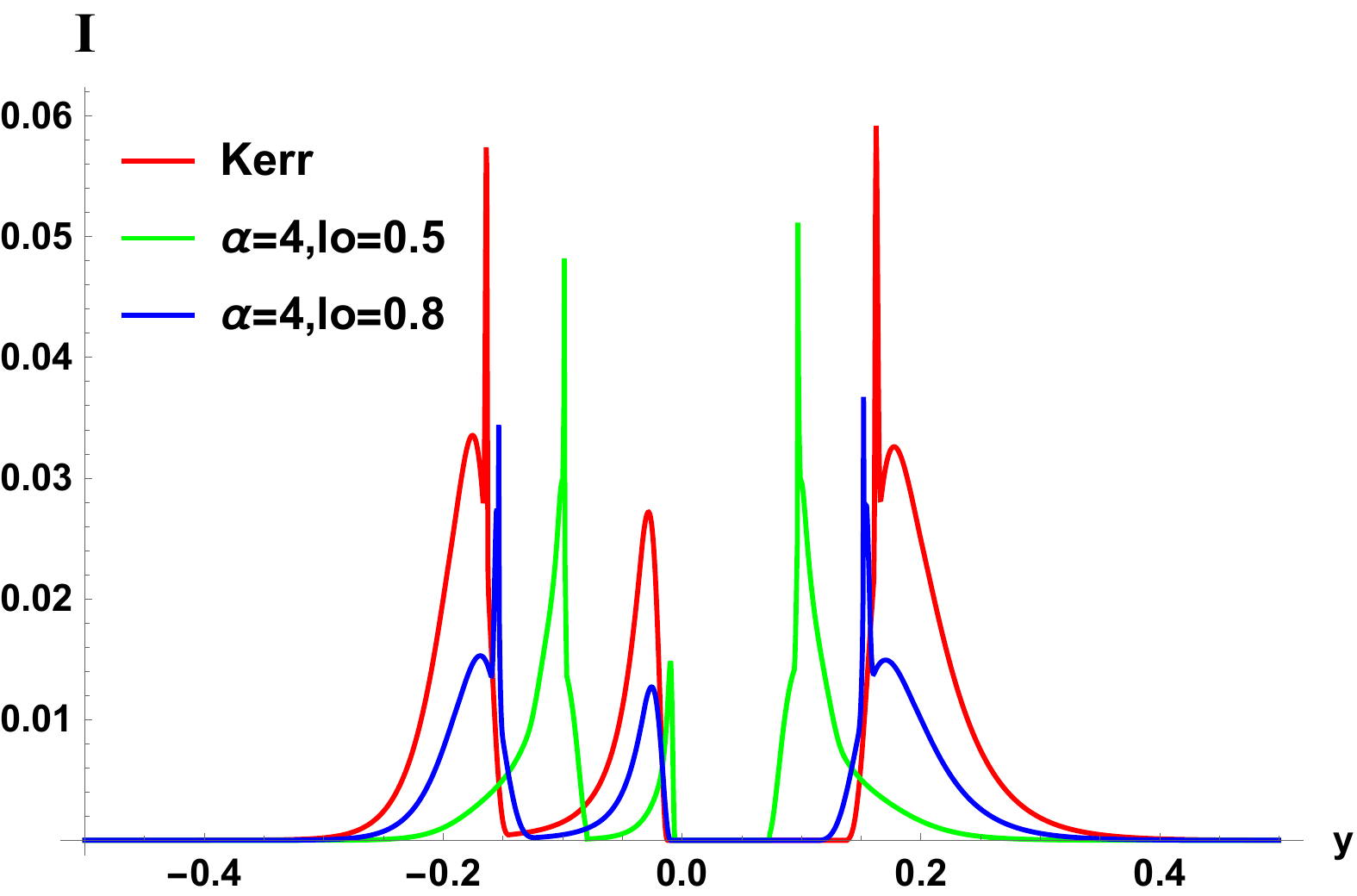}} \label{}\hspace{2mm}
\subfigure[\, $a=0.5$]
{\includegraphics[width=6cm]{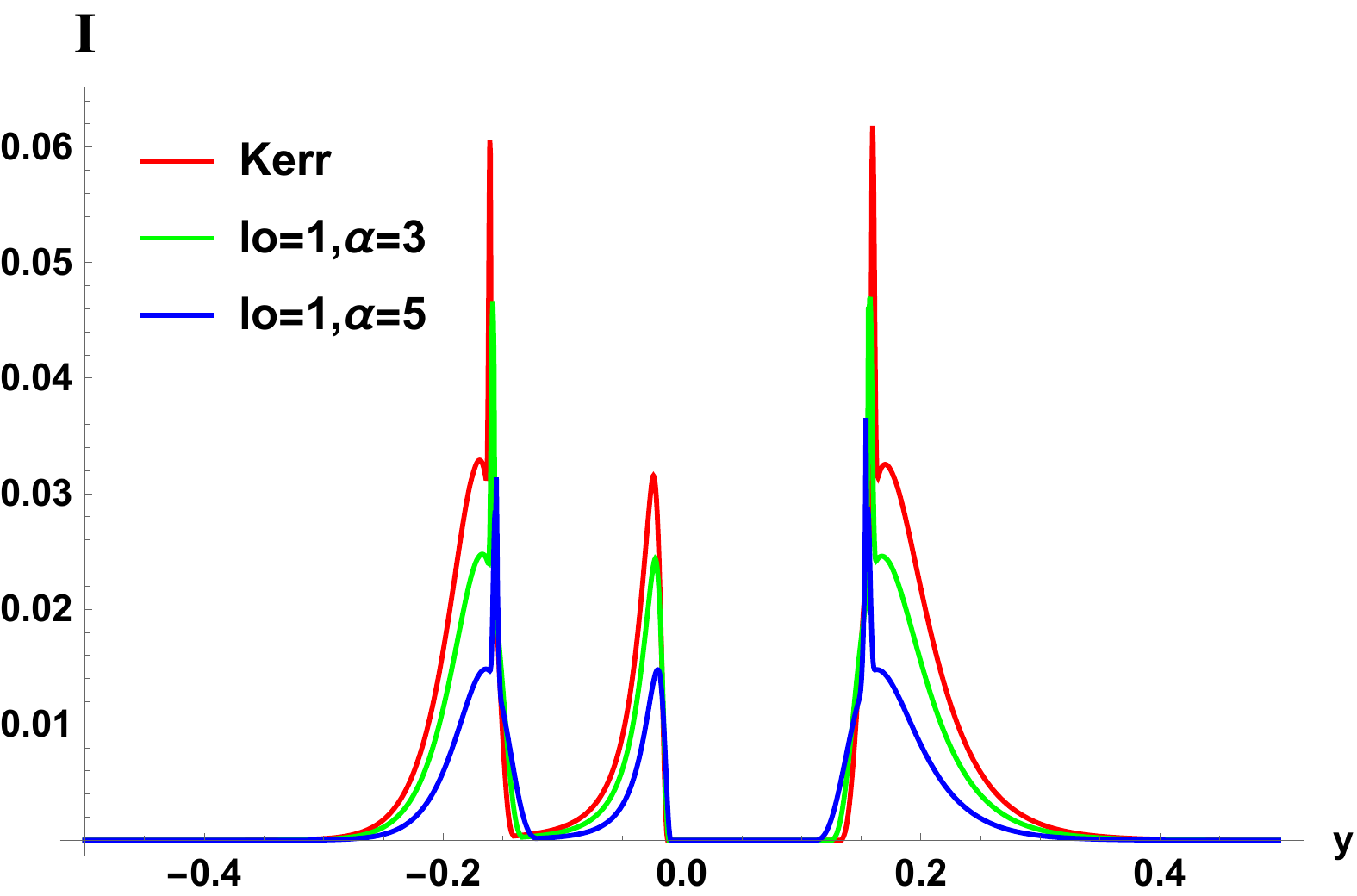}}\hspace{2mm}\\
\caption{The observed intensity distribution along $x-axis$ (upper panel) and $y-axis$ (bottom panel) for selected parameters. }
\label{fig-Intensity}
\end{figure}

In FIG. \ref{fig-Images}, we show the optical appearance of the (hairy) Kerr black hole illuminated by an accretion disk with prograde flow.  In each image,  we can see two universal features. One is  a shaded region at the center, the inner border of which wraps the dark region corresponding to direct emission image of  the event horizon, i.e.,  the inner shadow \cite{Chael:2021rjo}.  The other is a distinct bright photon ring which also {denotes} the critical curve of black hole shadow.
By comparing the subplots in each row with fixed spinning parameter, we find that both the inner shadow  and the critical curve  for the hairy Kerr black hole are smaller than those for Kerr black hole, which are similar for the feature of event horizon and ISCO addressed in last section.  In addition, as the hair parameter $l_o$ increases, both the inner shadow and critical curve increase. On the contrary, for a larger parameter $\alpha$, they become smaller. This means the effect of the competition between parameters $\alpha$ and $l_o$ on the size of the hairy black hole shadows which could have degeneracy comparing to GR, similar to the case with hairy  black hole \cite{Meng:2023htc,Meng:2024puu}.
We also observe the change in the brightness of the black hole image as $\alpha$ or $l_o$ shift. In order to clearly see this change,  we depict  the observed intensity \eqref{intensity-z} distribution of the (hairy) Kerr black hole along the $x-$axis and $y-$axis in FIG. \ref{fig-Intensity}. Due to the redshift effect, the observed intensity on the left side of the image is significantly greater than that on the right side, as shown in the upper panel. 


\begin{figure}[H]
\centering
\subfigure[\, $\text{Kerr},a=0.1$]
{\includegraphics[width=4.5cm]{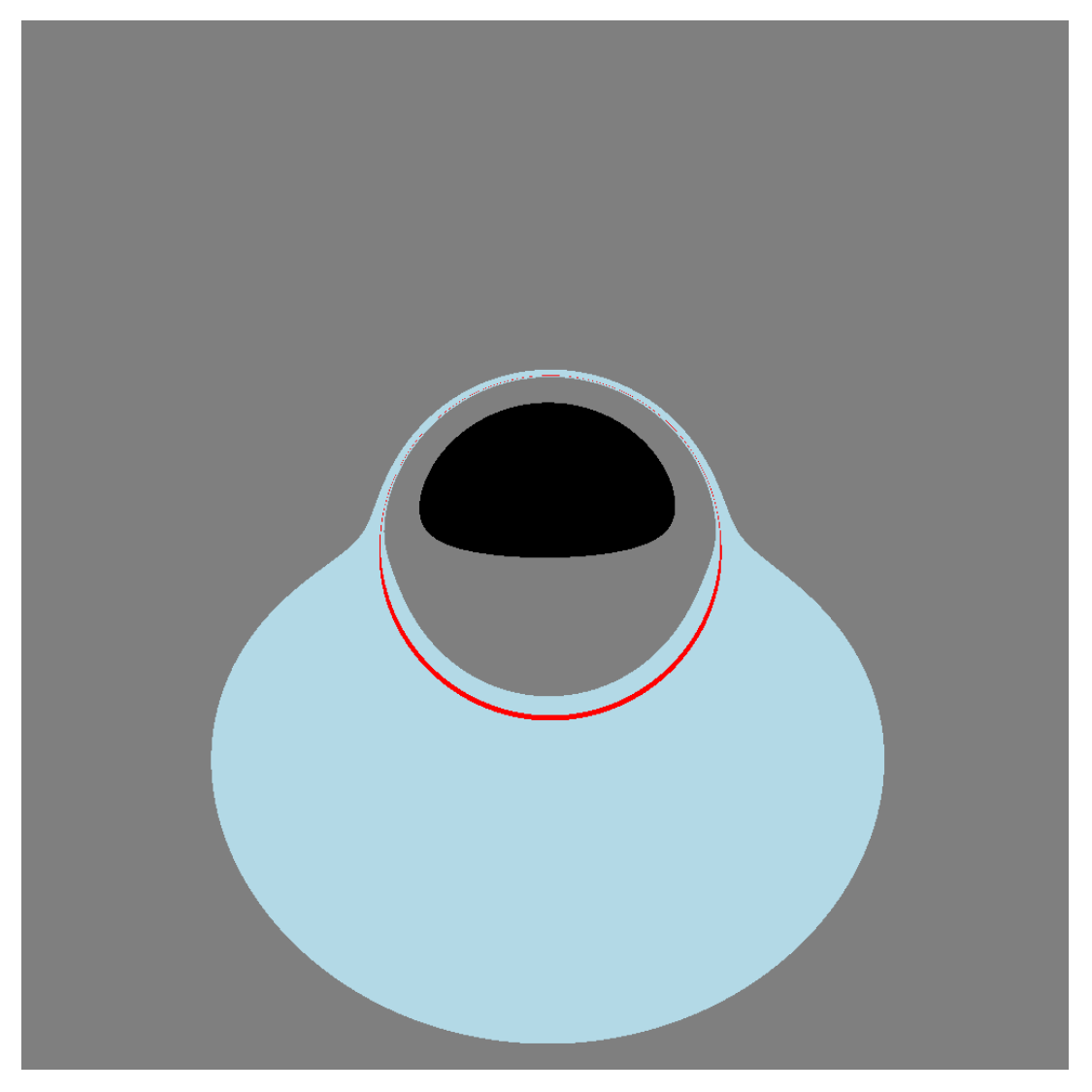}} \hspace{2mm}
\subfigure[\, $a=0.1,\alpha=4,l_o=0.8$]
{\includegraphics[width=4.5cm]{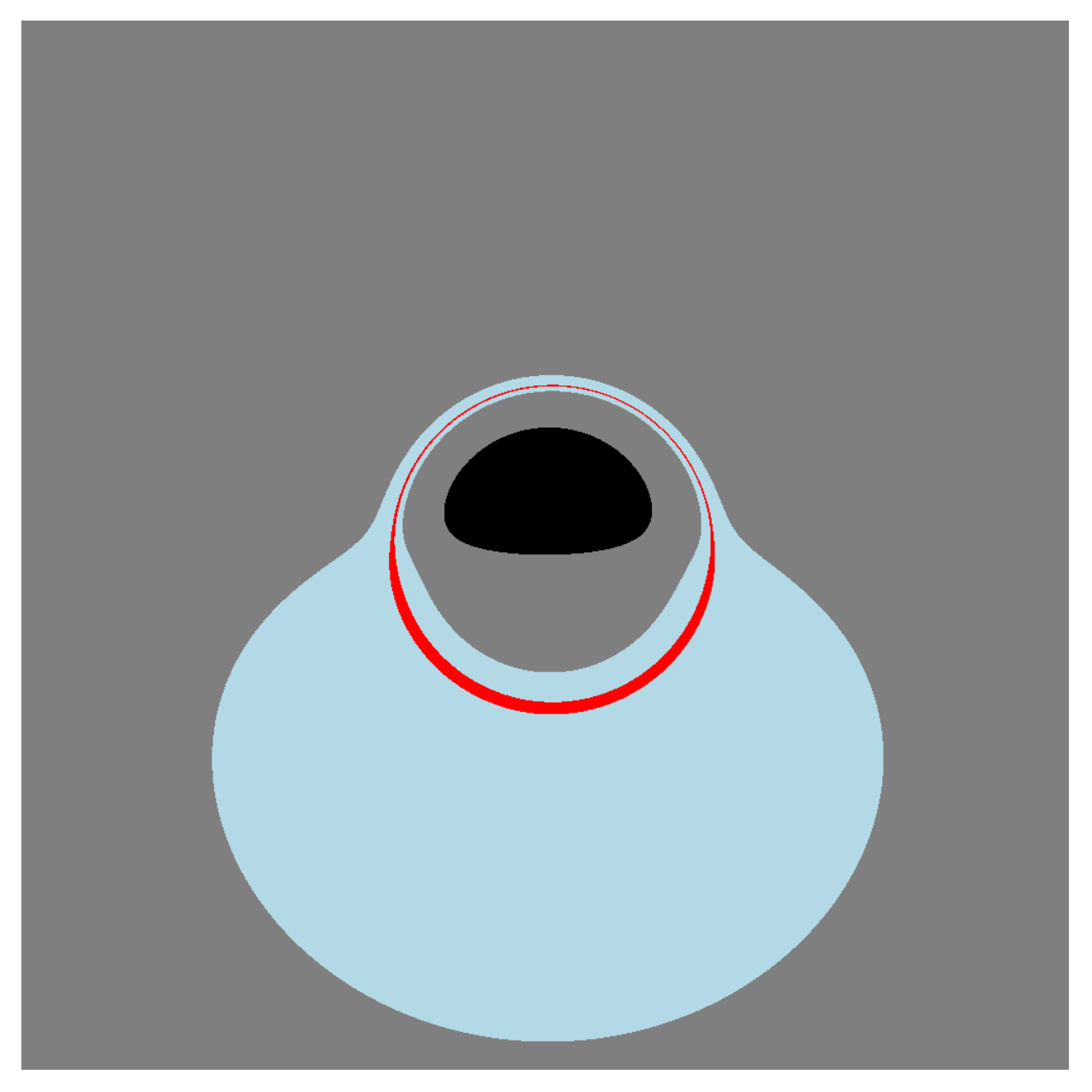}}\hspace{2mm}
\subfigure[\, $a=0.1,\alpha=4,l_o=0.5$]
{\includegraphics[width=4.5cm]{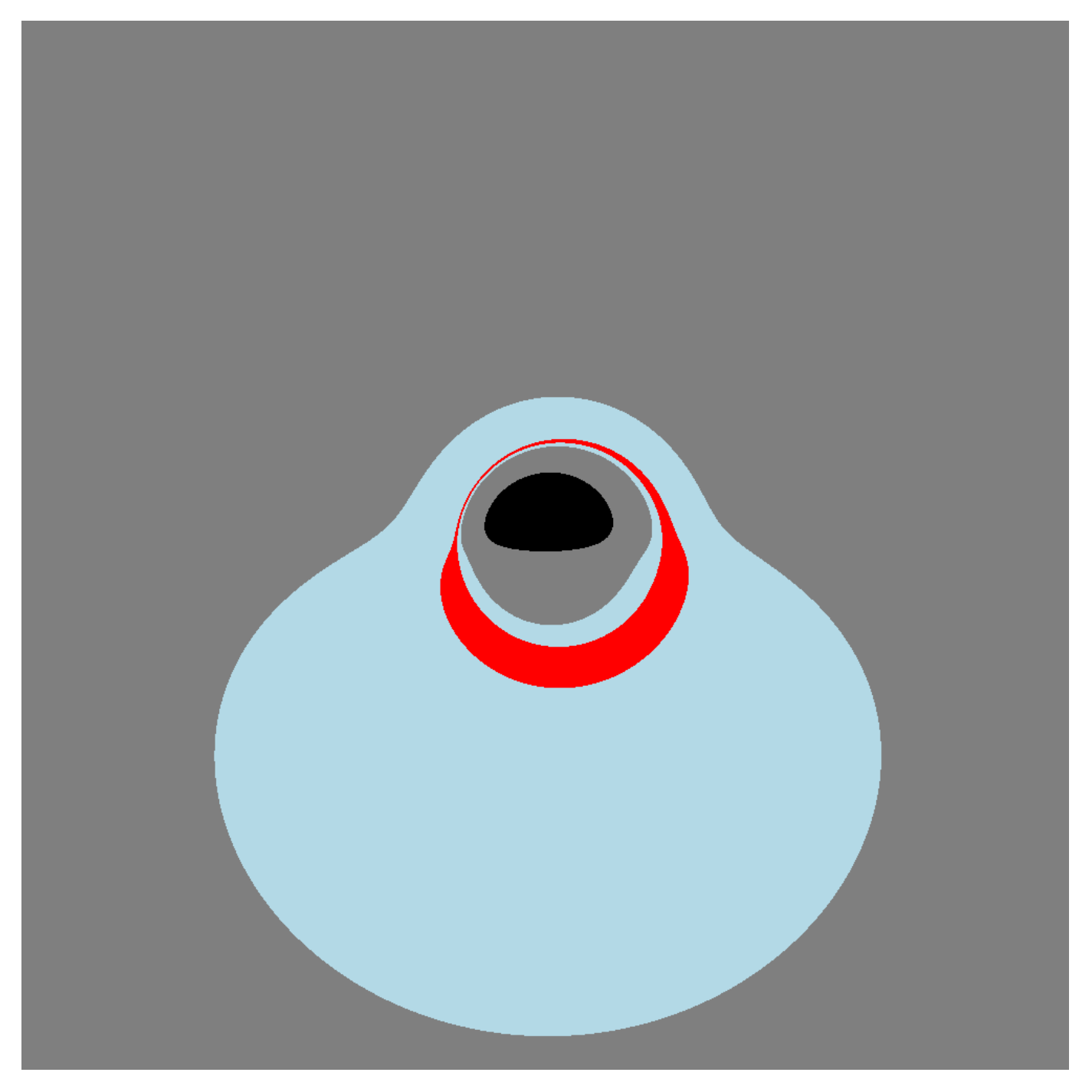}}\\
\subfigure[\, $\text{Kerr},a=0.5$]
{\includegraphics[width=4.5cm]{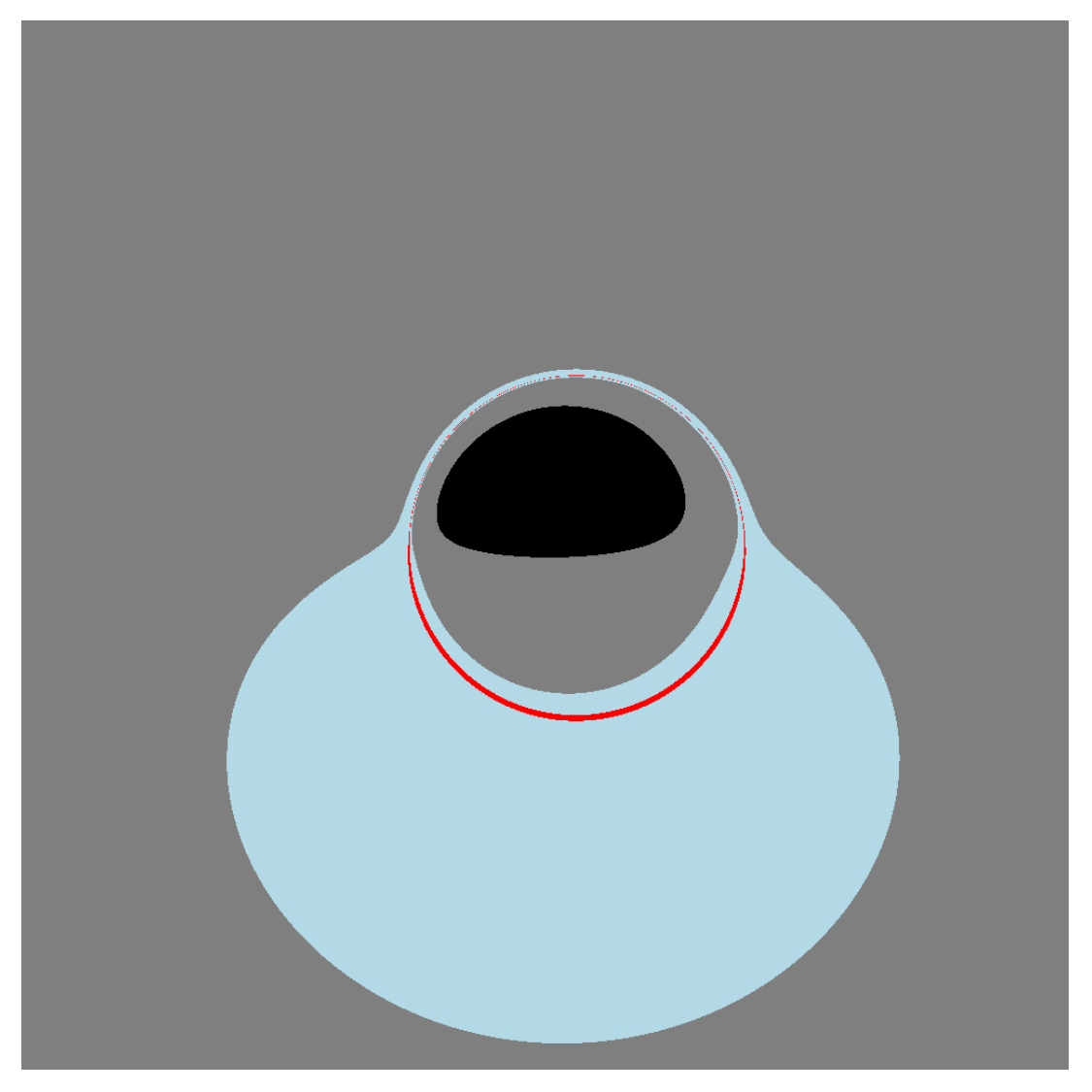}} \label{}\hspace{2mm}
\subfigure[\, $a=0.5,l_o=1,\alpha=3$]
{\includegraphics[width=4.5cm]{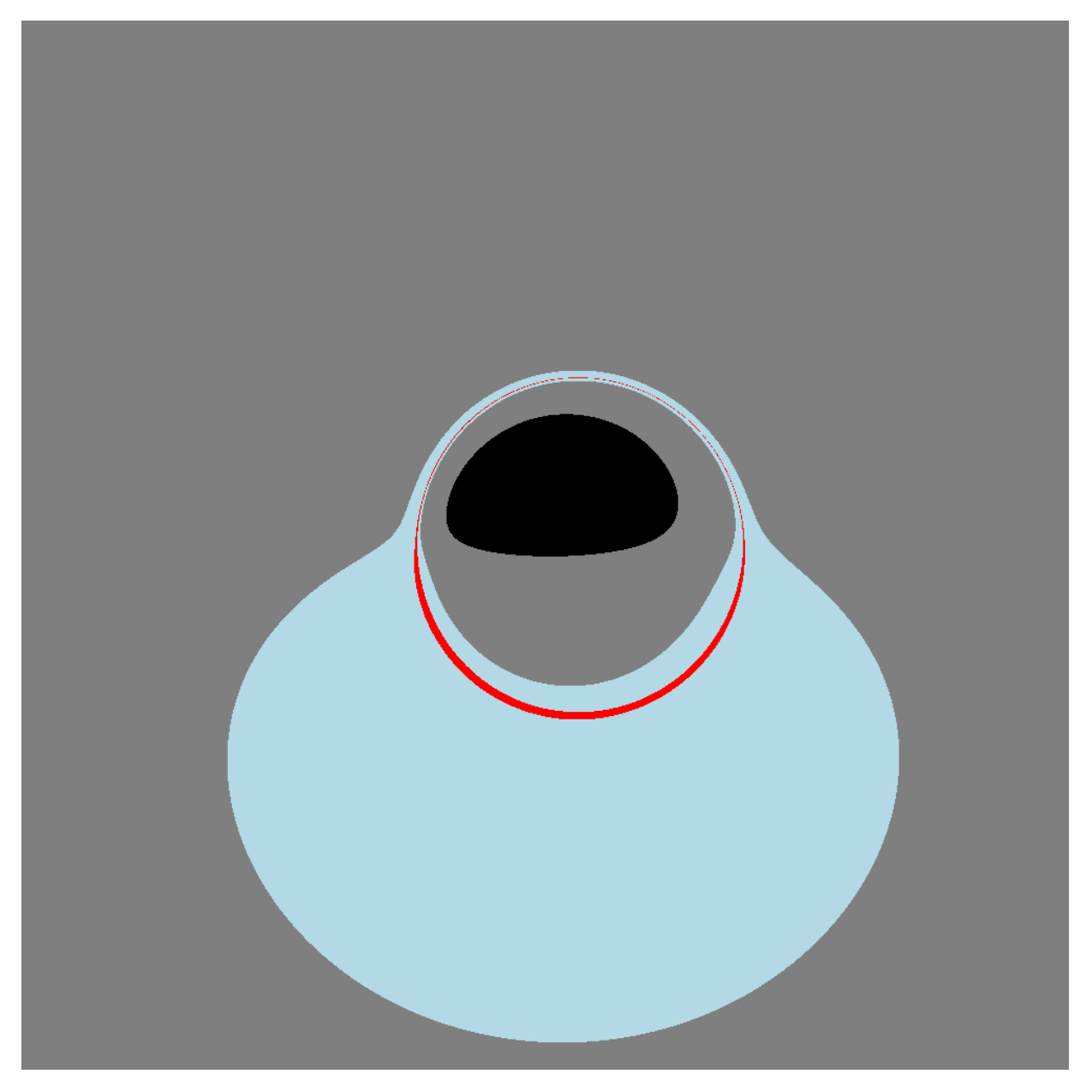}}\hspace{2mm}
\subfigure[\, $a=0.5,l_o=1,\alpha=5$]
{\includegraphics[width=4.5cm]{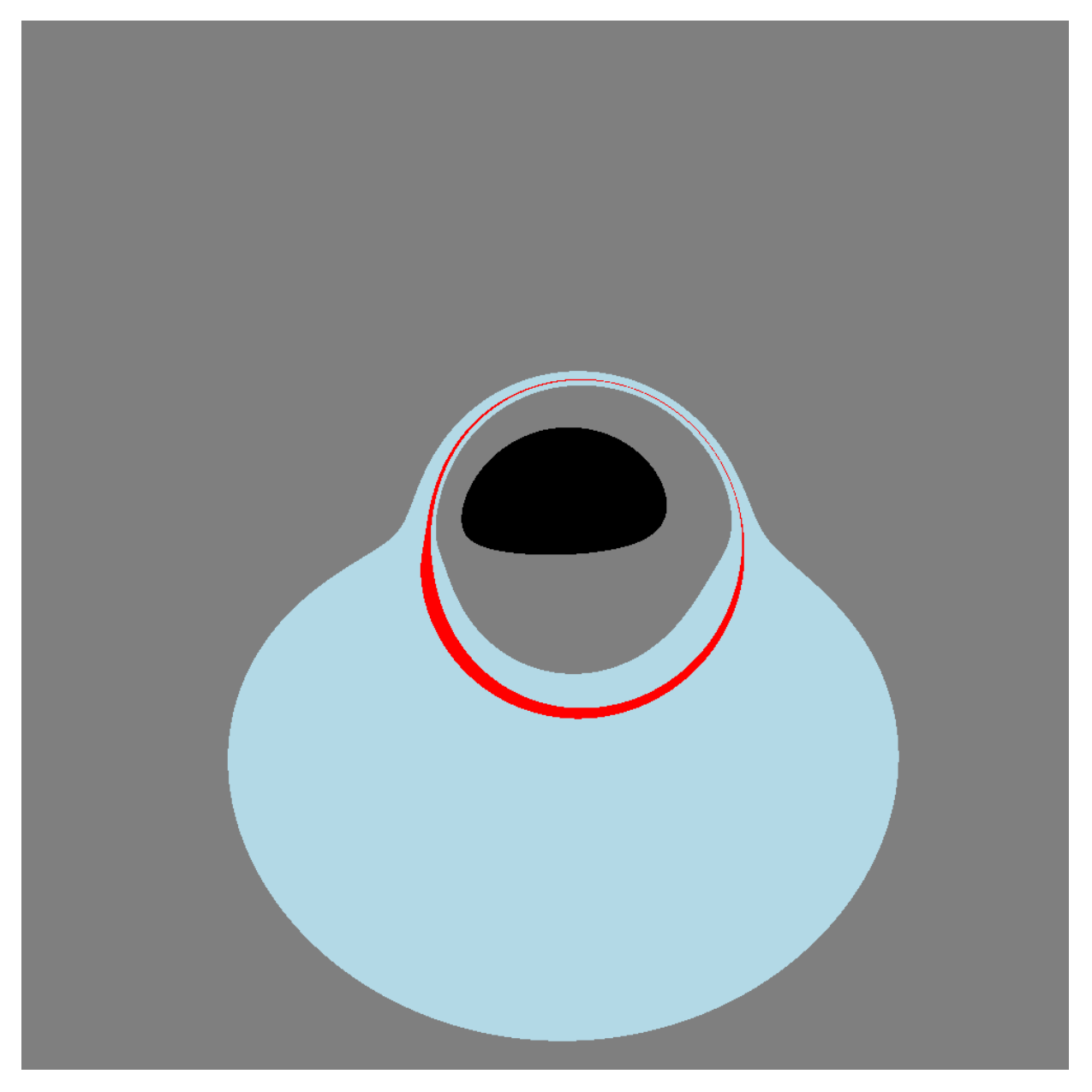}}\\
\caption{The images bands of three different emissions around the (hairy) Kerr black hole. The dark regions are the inner shadow. The gray, cyan and red band corresponds to the direct, lensed and photon ring emissions, respectively.}
\label{fig-emission}
\end{figure}

We then explicitly present the bands of three emissions of the accretion disk in FIG. \ref{fig-emission}.
In each plot, the central black region denotes the black hole;  the gray region represents the image band of direct emission of the disk, while the cyan and red regions indicate the images band of lensed and photon ring emission of the disk. Our results show that the effects of $\alpha$ and $l_o$ on the widths of direct emission and lensed ring emission are slight. However, for hairy Kerr black hole,  as the parameter $\alpha$ ($l_o$) increases (decreases), the photon ring in the image becomes more wider. This suggests that the photon rings of hairy Kerr black holes could be easier to be detected comparing against that of Kerr black hole.

\begin{figure}[H]
\centering
\subfigure[\, $\text{Kerr},a=0.1$]
{\includegraphics[width=4.5cm]{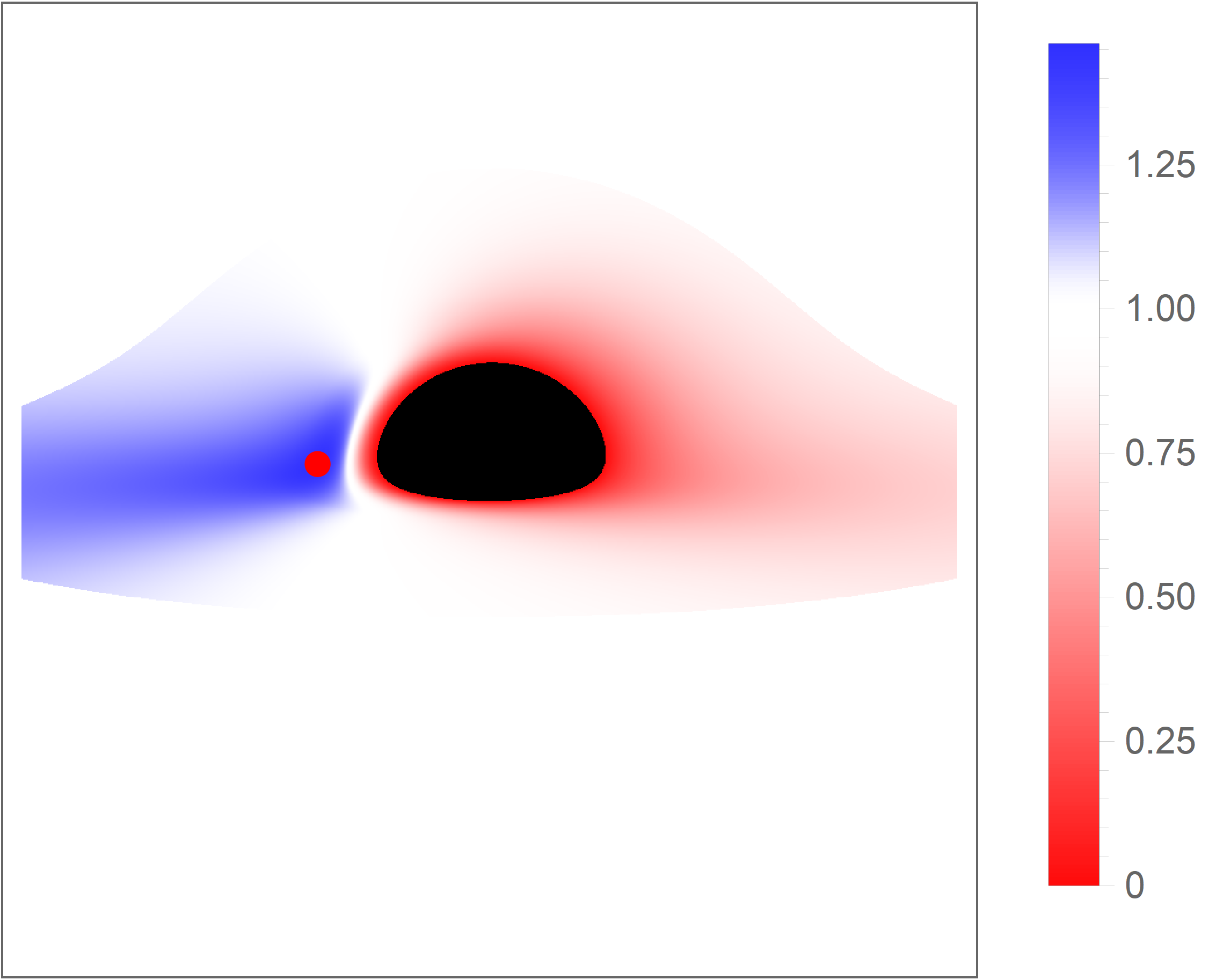}} \hspace{2mm}
\subfigure[\, $a=0.1,\alpha=4,l_o=0.8$]
{\includegraphics[width=4.5cm]{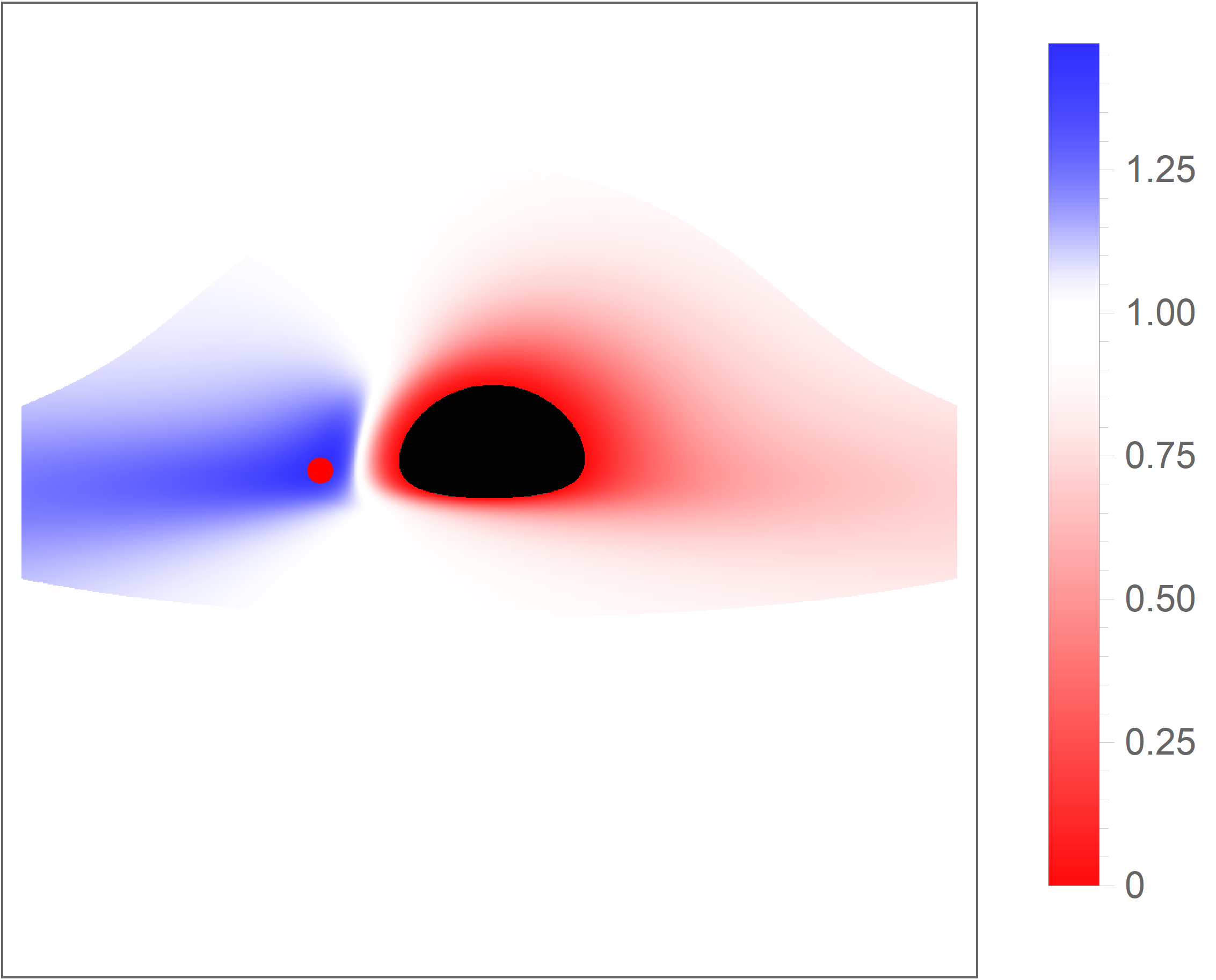}}\hspace{2mm}
\subfigure[\, $a=0.1,\alpha=4,l_o=0.5$]
{\includegraphics[width=4.5cm]{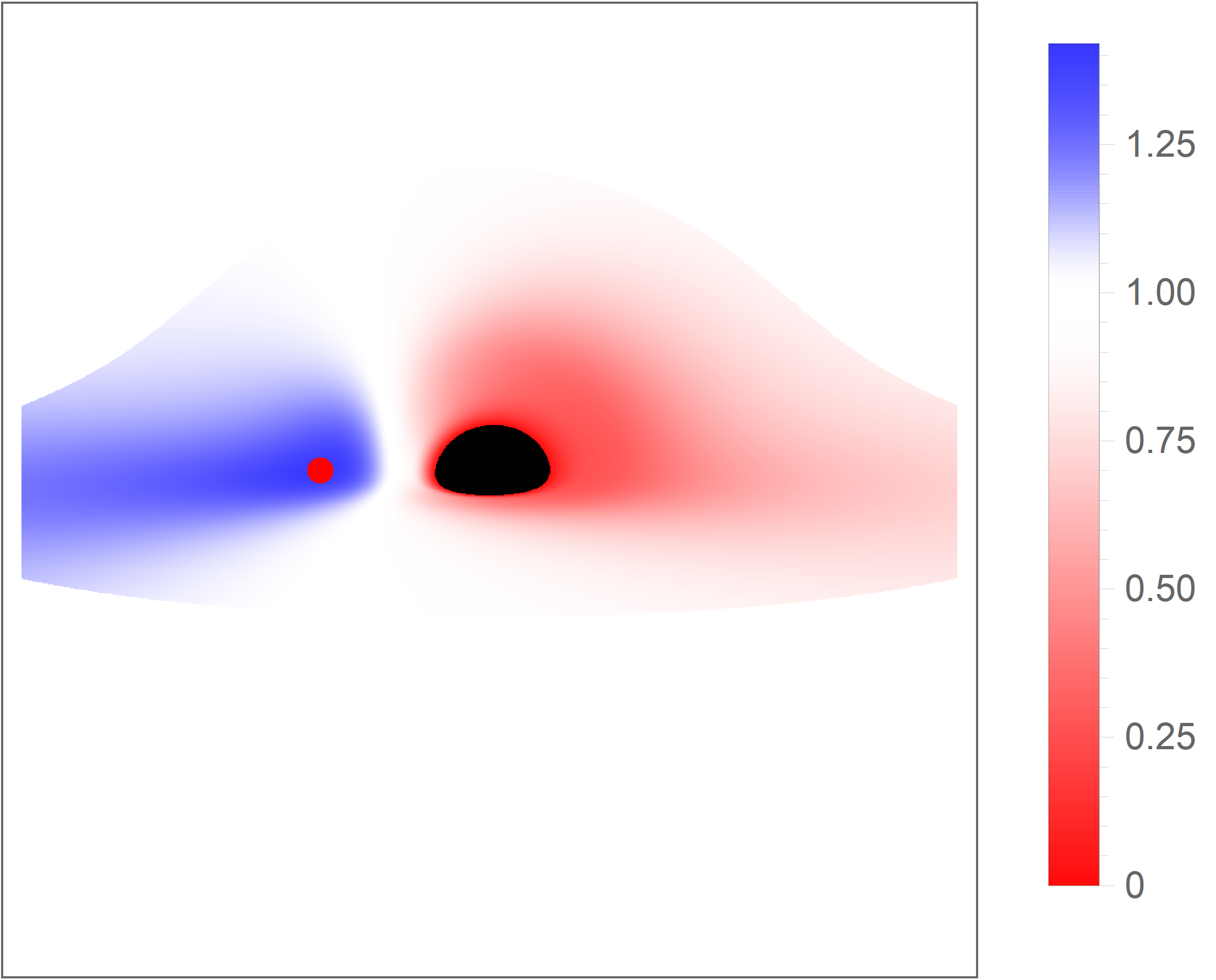}}\\
\subfigure[\, $\text{Kerr},a=0.5$]
{\includegraphics[width=4.5cm]{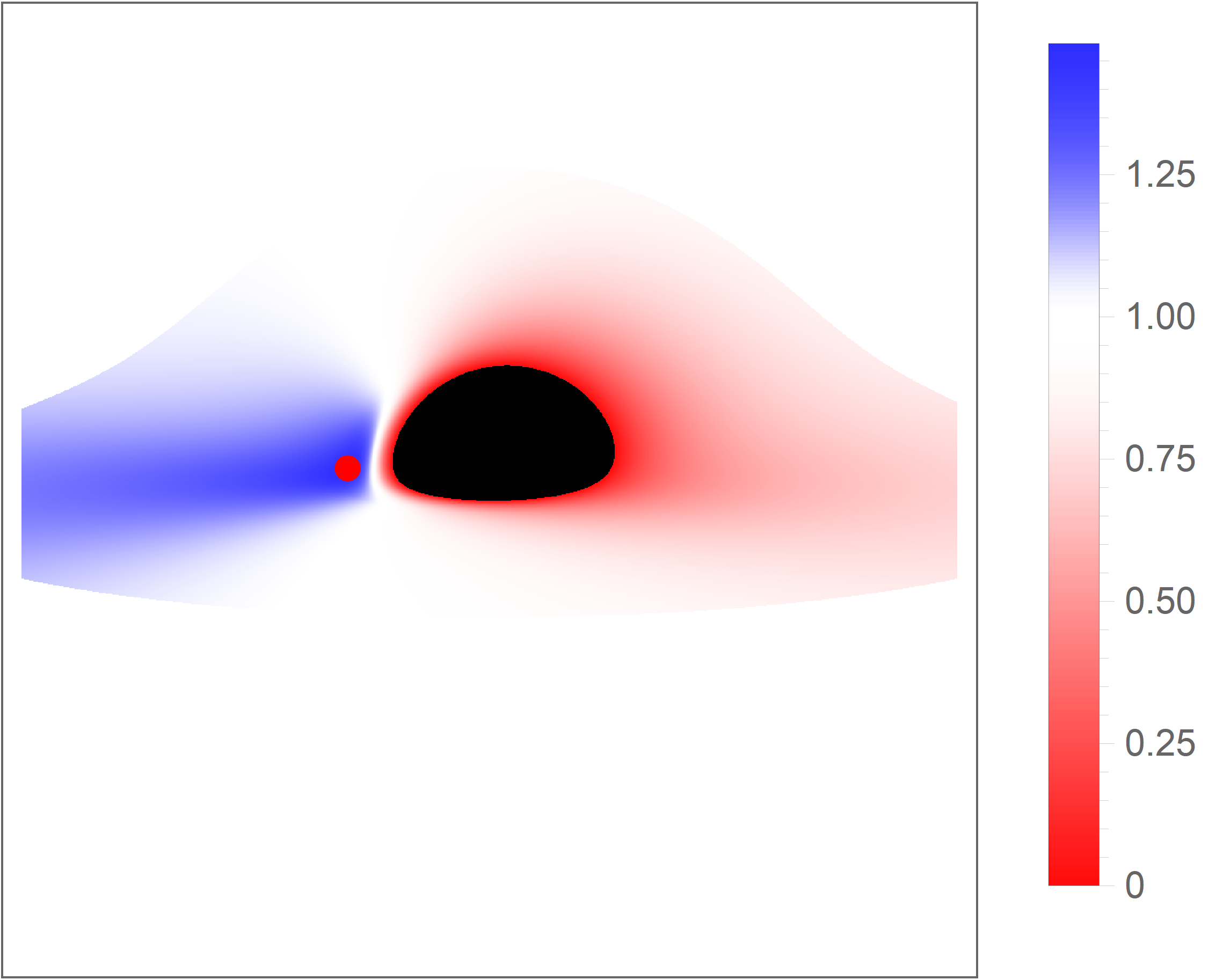}} \label{}\hspace{2mm}
\subfigure[\, $a=0.5,l_o=1,\alpha=3$]
{\includegraphics[width=4.5cm]{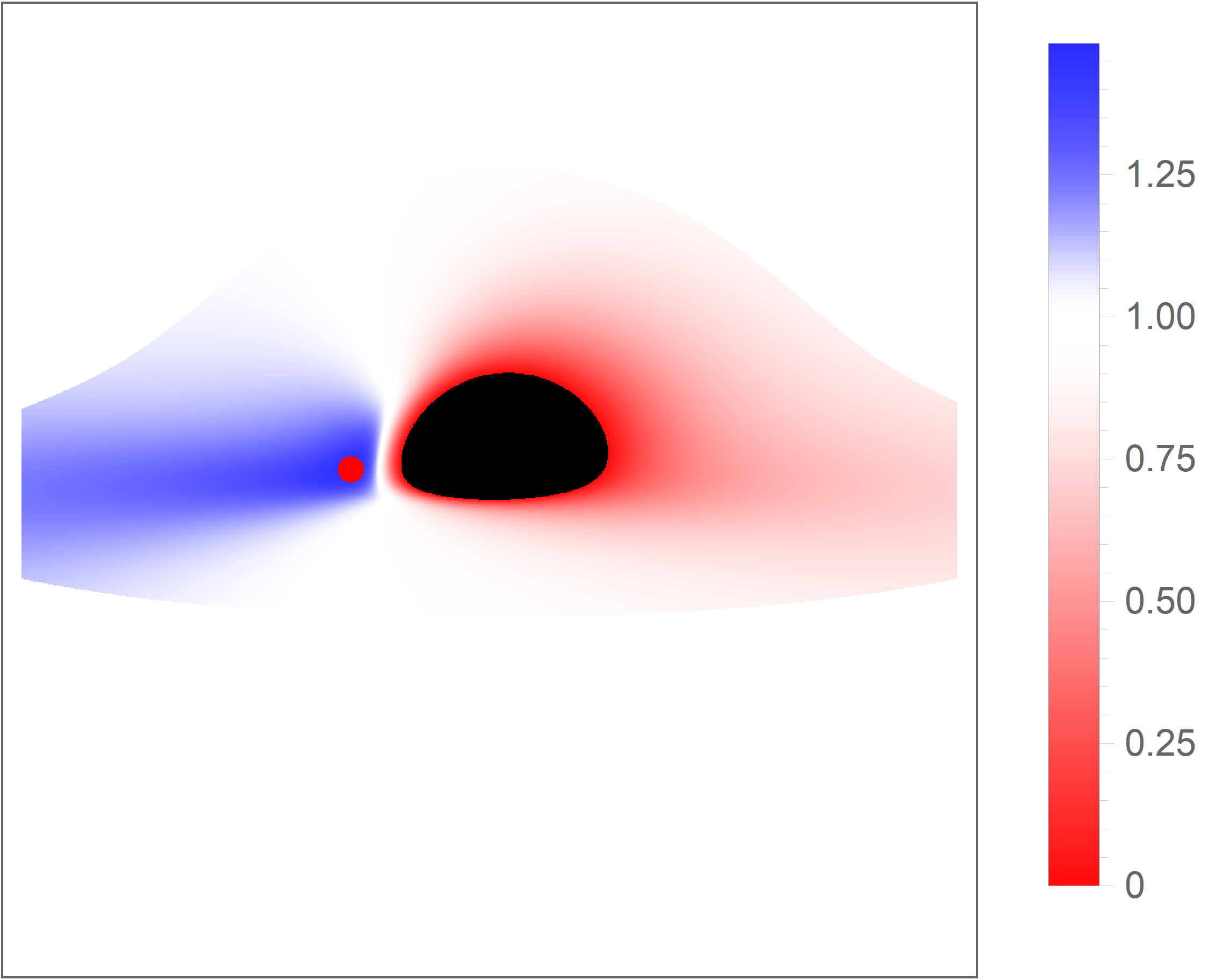}}\hspace{2mm}
\subfigure[\, $a=0.5,l_o=1,\alpha=5$]
{\includegraphics[width=4.5cm]{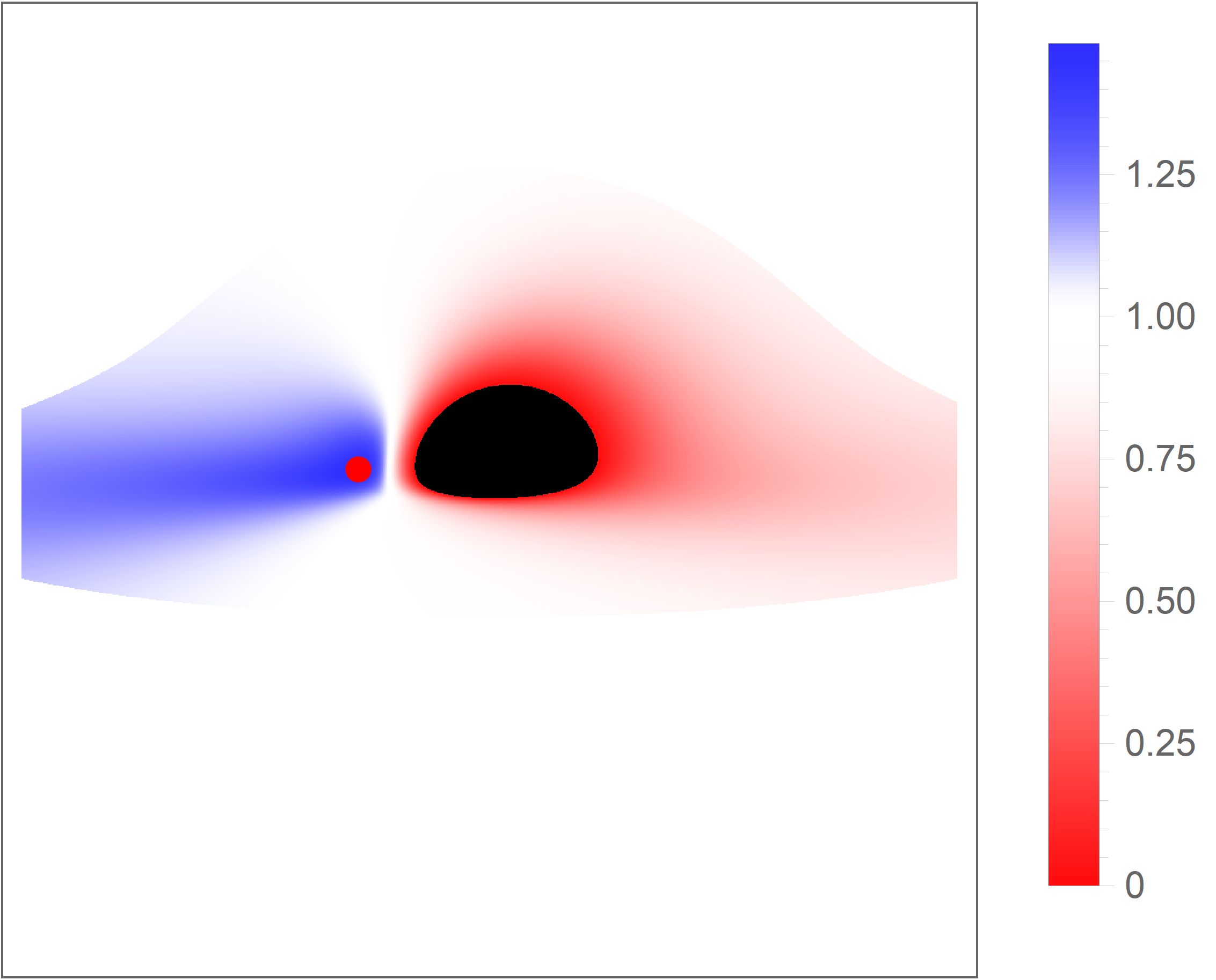}}\\
\caption{The redshift factor distributions of the direct images of the accretion disk. The red and blue represent redshift and blueshift, respectively.  And the red dot in each plot indicates the maximal blueshift point.}
\label{direct-redshift}
\end{figure}

\begin{figure}[H]
\centering
\subfigure[\, $\text{Kerr},a=0.1$]
{\includegraphics[width=4.5cm]{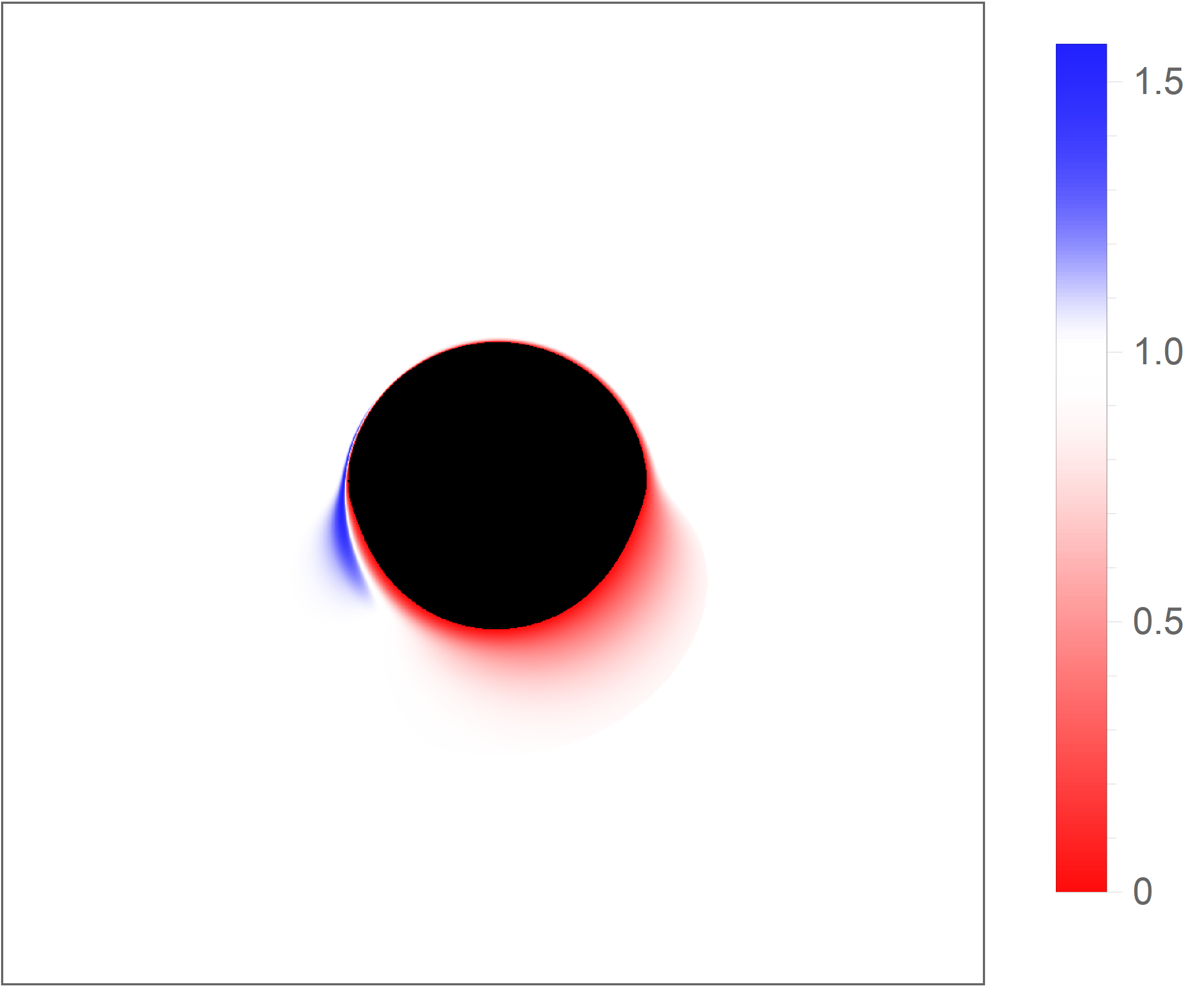}} \hspace{2mm}
\subfigure[\, $a=0.1,\alpha=4,l_o=0.8$]
{\includegraphics[width=4.5cm]{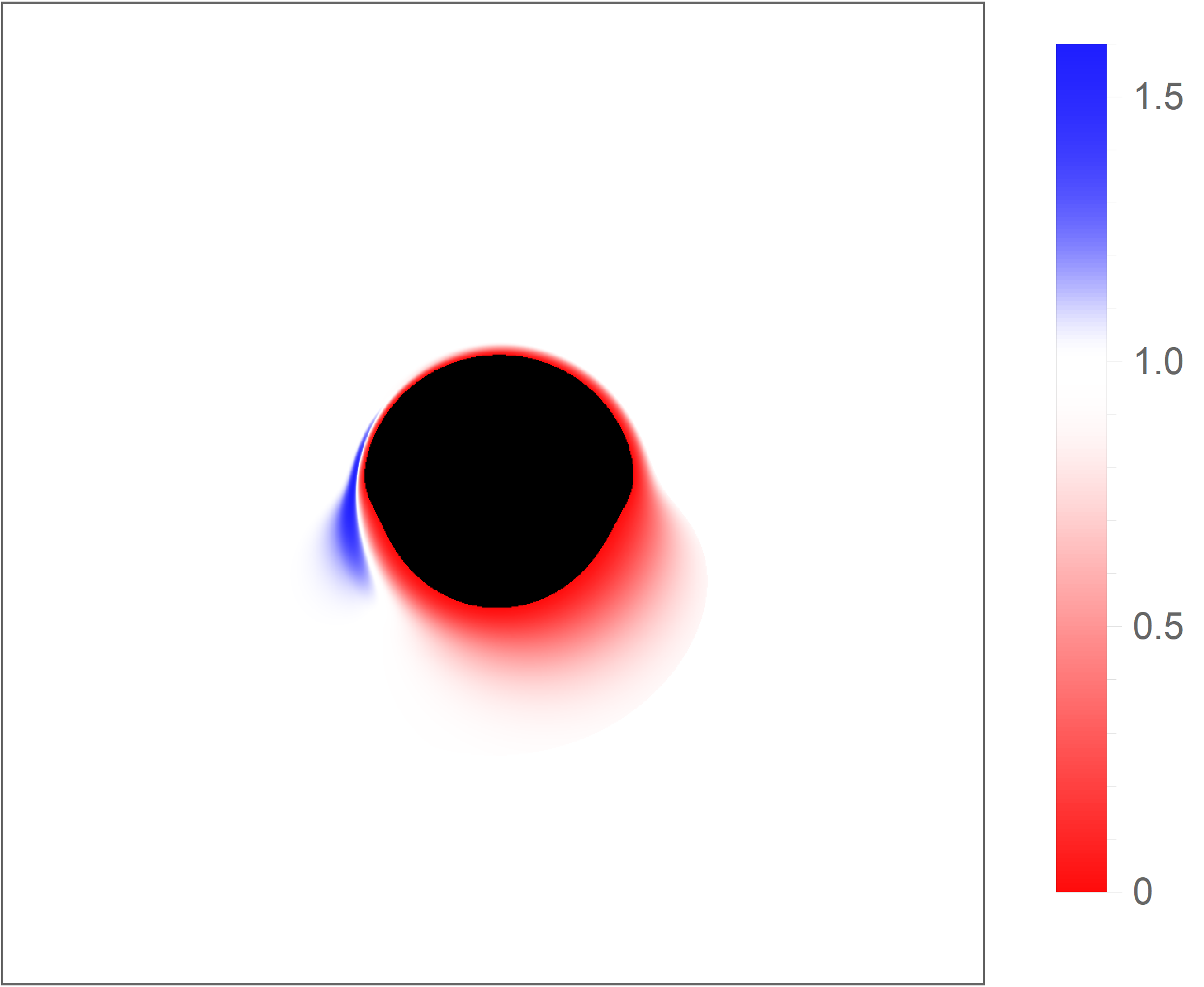}}\hspace{2mm}
\subfigure[\, $a=0.1,\alpha=4,l_o=0.5$]
{\includegraphics[width=4.5cm]{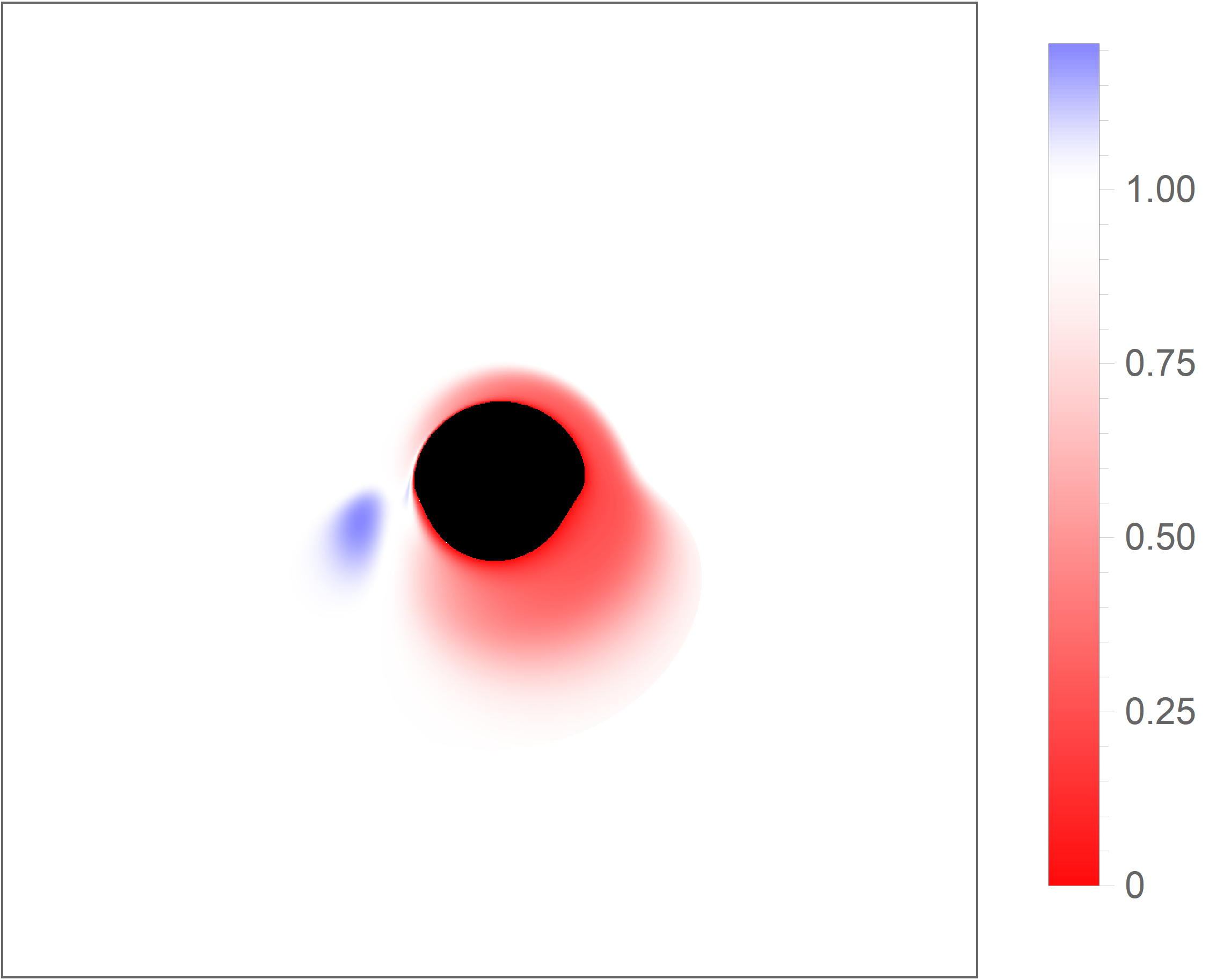}}\\
\subfigure[\, $\text{Kerr},a=0.5$]
{\includegraphics[width=4.5cm]{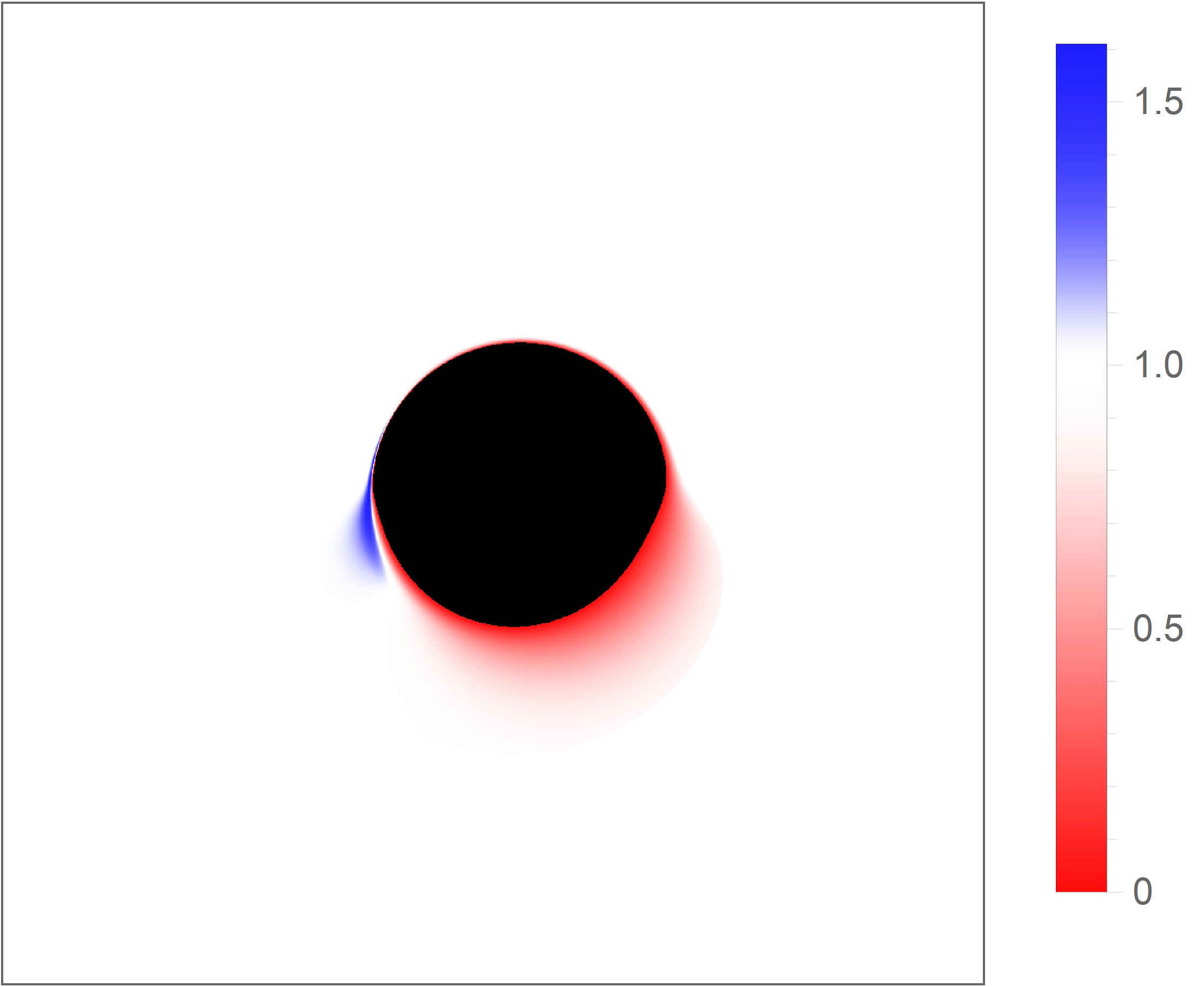}} \label{}\hspace{2mm}
\subfigure[\, $a=0.5,l_o=1,\alpha=3$]
{\includegraphics[width=4.5cm]{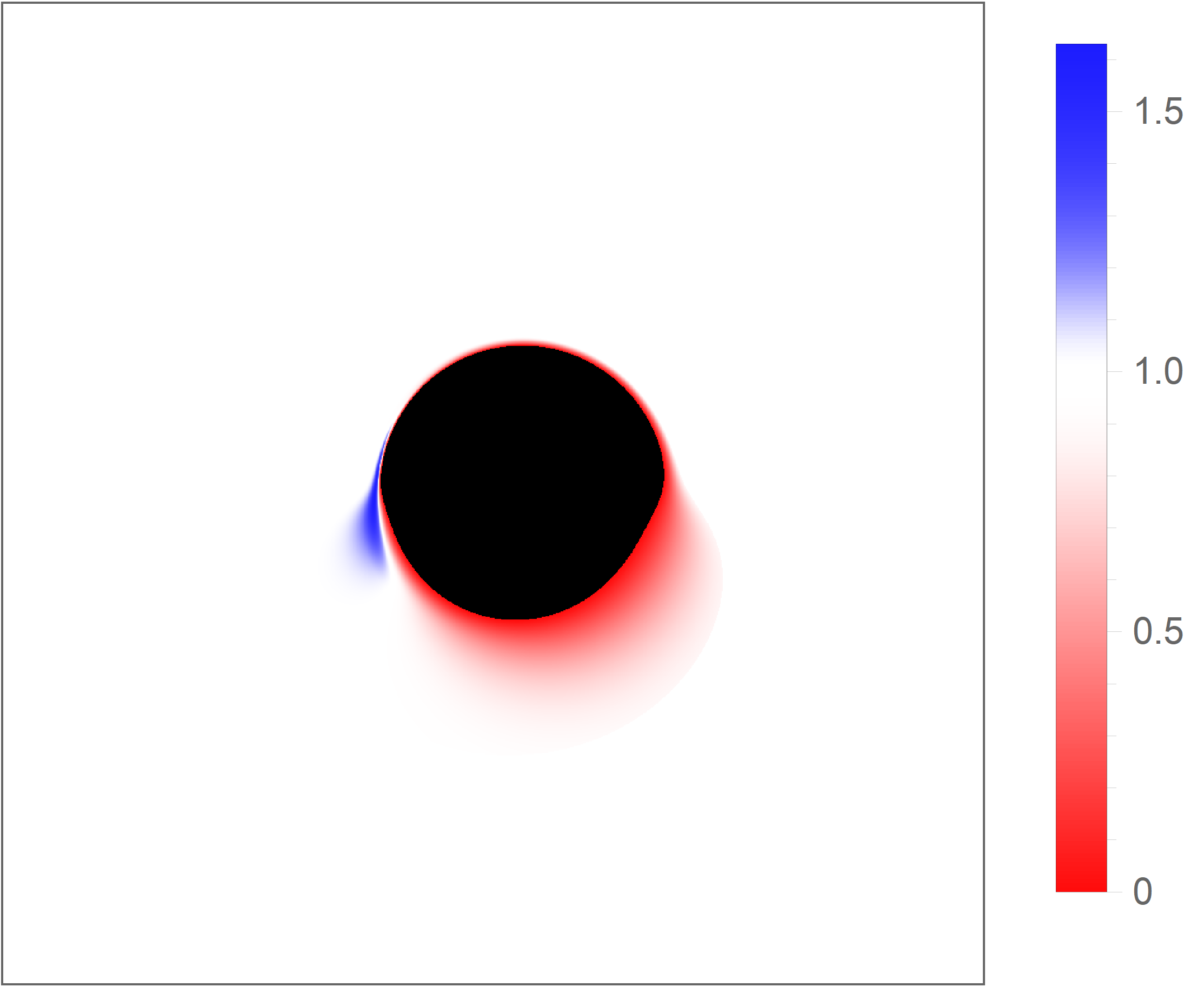}}\hspace{2mm}
\subfigure[\, $a=0.5,l_o=1,\alpha=5$]
{\includegraphics[width=4.5cm]{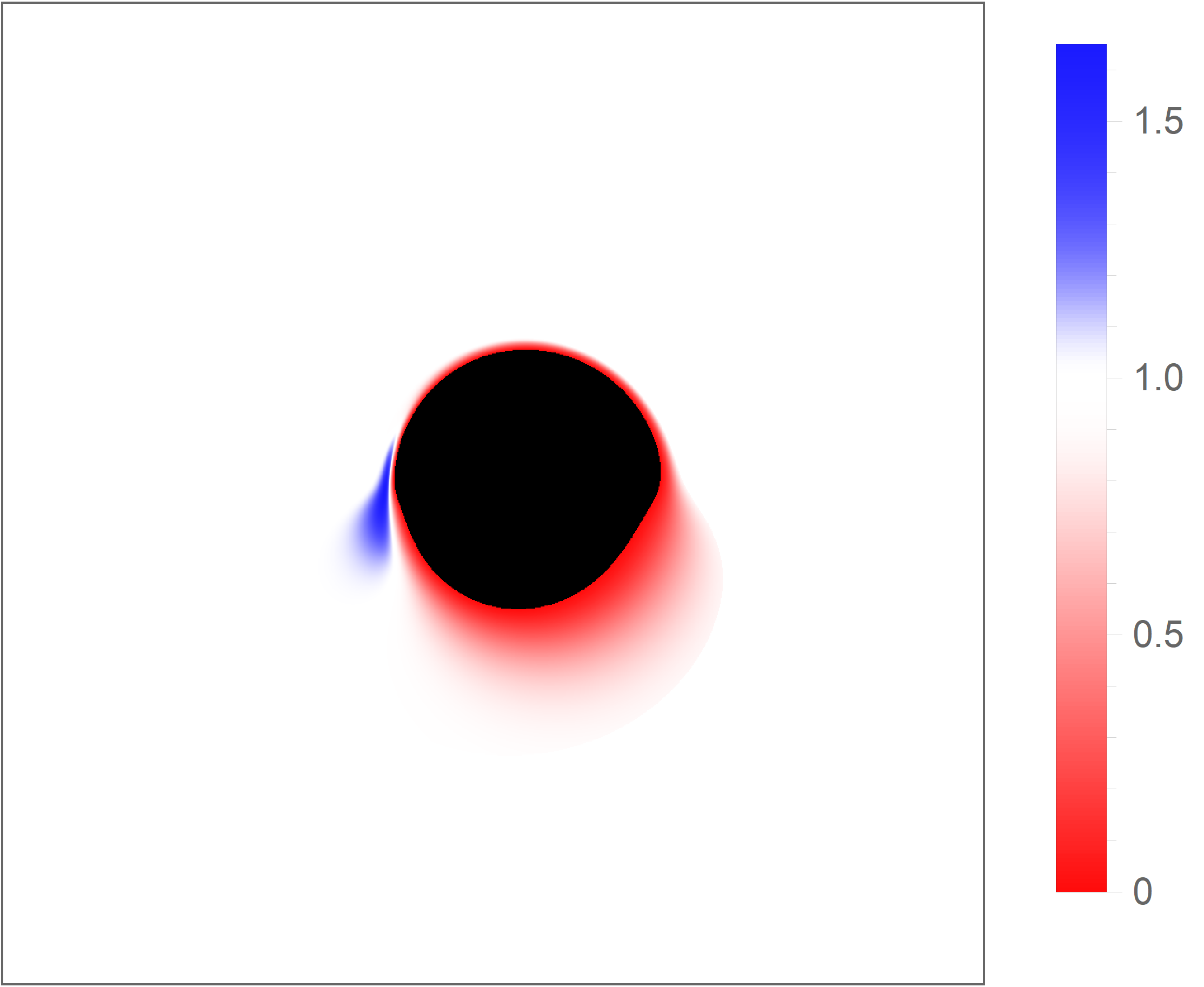}}\\
\caption{The redshift factors distribution of the lensed  images of the accretion disk. The red and blue represent redshift and blueshift, respectively.  And the border of the
black region in each plot gives  the lensed images of event horizon.}
\label{lensed-redshift}
\end{figure}

Furthermore, we show the redshifts of the direct and lensed images of the accretion disk in FIG. \ref{direct-redshift} and FIG. \ref{lensed-redshift}, respectively, in which  the red and blue represent redshift and blueshift  effects. From  FIG. \ref{direct-redshift} in which the black regions give the inner shadows,  the parameters $\alpha$ and $l_o$ have slight effects on the direct  images of the accretion disk, but we still can see that the maximal blueshift point indicated by red dot are shifted by the change of the parameters. From FIG. \ref{lensed-redshift} in which the edges of the black regions are the lensed images of the event horizon, we see that the blueshift could be enhanced by the increasing of $\alpha$ and $l_o$, though they have competitive effects on the size of the black region. 

\section{Conclusion and discussion}\label{conclusion}

The black hole images released by the EHT collaborators open a new window for testing gravity in strong field regime and also provide a new exploration opportunity for verifying the violations or validity of the no-hair theorem. In our previous works we explored the shadows, rings and optical images of the hairy Schwarzschild and  Reissner-Nordstr\"{o}m black holes both in the scenario of GD proposal \cite{Meng:2023htc,Meng:2024puu}, suggesting that the hairy parameters could significantly influence the distributions of the photon trajectories. We also found that the optical appearance between the hairy and no-hair black holes could have degeneracies. To extend these results into rotating case and check their universality, in this paper we investigated the optical appearance of a hairy Kerr black hole illuminated by a thin accretion disk and further explored the imprint of the hair in the black hole image.

First, we investigated the event horizon of the hairy Kerr black hole and the ISCOs of the massive neutral particles. As the parameter $\alpha$ increases, the event horizon of the black hole becomes smaller. Conversely, for a larger principal primary parameter $l_o$, the event horizon of the black hole is also larger. It is not difficult to deduce that the ISCOs of the massive particle vary with the parameters $\alpha$ and $l_o$ in a manner similar to the event horizon of the hairy Kerr black hole. Then we review the setup and methods for imaging the black hole with an accretion disk on the equatorial plane. Obviously, due to the form of deformed term in the metric one can expect that for appropriate parameters $\alpha$ and $l_o$, the image of the hairy Kerr black hole could be similar to that of the Kerr black hole. However, for the general parameters $\alpha$ and $l_o$ the images of the hairy Kerr black hole and the Kerr black hole become significantly different, which makes it possible for us to distinguish between the two in astronomical observations. It is worth noting that as the parameter $l_o$ decreases, the shadow size of the black hole becomes smaller, while as the parameter $\alpha$ increases, the shadow size decreases.

Next, we investigated the widths of the direct, lensed ring and photon ring emission regions for the thin accretion disk. The widths of the direct emission and lensed ring emission change slightly with the parameters $\alpha$ and $l_o$, whereas the width of the photon ring emission changes significantly. Surprisingly, as the primary hair parameter $l_o$ decreases, the width of the photon ring emission increases dramatically. This behavior theoretically indicates that it may be easier to observe the photon ring of the hairy Kerr black hole than that of Kerr black hole. Finally, we investigated the redshift of the direct and lensed images, showing that the parameters have slight effects on the direct images of the accretion disk but the maximal blueshift point is shifted by the change of the parameters. Besides, the blueshift of the lensed image become smaller for smaller parameters $\alpha$ and $l_o$.

In conclusion, we investigated the shadows and images of hairy Kerr black holes illuminated by a thin accretion disk. For appropriate parameters $\alpha$ and $l_o$, the image of the hairy Kerr black hole is significantly different from that of the Kerr black hole, which may provide an attempt for us to check whether the Kerr black hole is hairy from astronomical observations. Moreover, $\alpha$ and $l_o$ have competing effects on the shadow size of the hairy Kerr black hole, so the hairy parameters can be determined from the image of the hairy Kerr black hole. In addition, the hairy parameters significantly increase the region of the photon ring emission, making it a better choice to observe the photon rings of the hairy Kerr black hole in astronomical observations.

\begin{acknowledgments}
We appreciate Minyong Guo for his indispensable  help  in numeric. This work is partly supported by the Natural Science Foundation of China under Grants No. 12375054 and 12447137,  and the Postgraduate Research $\&$ Practice Innovation Program of Jiangsu Province under Grants No. KYCX22$_{-}$3452. YZ is also supported by the research funds No. 2055072312.

{\bf Note added}: this version was submitted into European Physical Journal C. During the review processing, we read the nice paper  in arXiv \cite{Li:2025ixk}, in which the authors also studied the radiative property and optical appearance of a rotating hairy black hole surrounded by a thin accretion disk. It is noted that the inner side of their accretion disk model is the ISCO of the black hoke, differentiating from ours.
\end{acknowledgments}

\bibliography{ref}
\bibliographystyle{utphys}

\end{document}